\documentclass[a4paper,fleqn,usenatbib]{mnras}
\usepackage[dvipsnames,usenames]{color}
\usepackage{colortbl}
\usepackage{mathptmx}
\usepackage{etoolbox}
\usepackage{graphicx}
\usepackage{amsmath}
\usepackage{natbib}
\newcommand{\etal}{{et al}\/.}
\newcommand\rd{{\rm d}}
\pubyear{2017}
\newtoggle{highq}
\togglefalse{highq}
\begin{document}
\title[A model of radio galaxies]{A simulation-based analytic model of radio galaxies}
\author[M.J.\ Hardcastle]
{M.J.\ Hardcastle\\
Centre for Astrophysics Research, School of Physics, Astronomy and Mathematics, University of
  Hertfordshire, College Lane, Hatfield AL10 9AB, UK\\
}
\maketitle
\begin{abstract}
I derive and discuss a simple semi-analytical model of the evolution
of powerful radio galaxies which is not based on assumptions of
self-similar growth, but rather implements some insights about the
dynamics and energetics of these systems derived from numerical
simulations, and can be applied to arbitrary pressure/density profiles
of the host environment. The model can qualitatively and
quantitatively reproduce the source dynamics and synchrotron light
curves derived from numerical modelling. Approximate corrections for
radiative and adiabatic losses allow it to predict the evolution of
radio spectral index and of inverse-Compton emission both for active
sources and for `remnant' sources after the jet has turned off. Code
to implement the model is publicly available. Using a standard model
with a light relativistic (electron-positron) jet, sub-equipartition
magnetic fields, and a range of realistic group/cluster environments,
I simulate populations of sources and show that the model can
reproduce the range of properties of powerful radio sources as well as
observed trends in the relationship between jet power and radio
luminosity, and predicts their dependence on redshift and environment.
I show that the distribution of source lifetimes has a significant
effect on both the source length distribution and the fraction of
remnant sources expected in observations, and so can in principle be
constrained by observations. The remnant fraction is expected to be
low even at low redshift and low observing frequency due to the rapid
luminosity evolution of remnants, and to tend rapidly to zero at high
redshift due to inverse-Compton losses.
\end{abstract}

\section{Introduction}
\label{sec:intro}

\subsection{The evolution of radio sources}

A key problem in the study of radio-loud active galaxies (RLAGN) is the
difficulty of inferring their dynamical state from observation. The
large-scale structures we see are dynamic, not static; the size, shape
and luminosity of the lobes of RLAGN change with time. A consequence
of this fact is that observational selection criteria imply
biases that need to be accounted for in interpretation.

The basic nature of radio source dynamics has been understood since at
least \cite{Scheuer74}, but is complex in detail. Let's first consider
a source that starts in a spherically symmetric environment, at $t=0$ and continues
with a constant two-sided jet power $Q$ (determined by the accretion
system) until the jet turns off at time $t=T$. I denote the total source radio
luminosity at some fixed observing frequency as $L$ and the lobe length as $R$. Clearly, at $t=0$, $L=0$
and $R=0$. At later times $t<T$, the growth of the lobe ($\rd R/\rd
t$) is determined by the balance of the jet momentum flux and the
internal lobe pressure, on the one hand, against the external thermal
pressure and the ram pressure due to the lobe's expansion on the
other. If jets are light and relativistic when they are generated, then the momentum
flux of each jet is simply $Q/2c$, and the growth of the source is
determined solely by the interaction with the environment with no free
parameters. Unless the internal pressure in the lobe falls below the
external thermal pressure, or there are catastrophic interactions with
the external environment not modelled in this picture, such as
high-speed bulk motions, $\rd R/\rd t > 0$ at all times. A source's
size is thus a proxy of its age, but also its environment -- without a
knowledge of environment a physical size only gives a weak lower limit
on age.

The radio luminosity (density) of such a source is dependent on the
energy density of radiating particles and field in the lobe, and on
the lobe volume, so in general initially we expect luminosity to rise
with time ($\rd L/\rd t > 0$), though this may be masked at a given
observing frequency by effects such as self-absorption for small
sources. Numerical models
\citep{Hardcastle+Krause13,Hardcastle+Krause14} predict that the radio
luminosity will then enter a phase where it is roughly constant with
time (the lobe is still growing, but its pressure is falling) before
starting to drop off again at late times. The peak radio luminosity
for a given jet power will depend on environment
\citep{Barthel+Arnaud96,Hardcastle+Krause13}. The effect of radiative
losses (`spectral ageing') will start to become important at a time
that depends on the observing frequency, which adds an additional
time-dependent declining term to the radio luminosity. So the overall
expected picture is one of a relatively rapid rise, peaking at a
physical size comparable to the radius at which the pressure starts to
drop off rapidly with distance, followed by gradual decline. After the
jet turns off, at $t>T$, the decline is expected to be more rapid as
the contents of the lobes are no longer replenished with newly
accelerated electrons. Analytic models agree with this general
picture, though with differences in detail depending on the
assumptions made \citep[e.g.,][]{Kaiser+Best07}.

Even without making this picture more quantitative there are some
extremely important implications:
\begin{enumerate}
  \item Radio luminosity is a function of time, not just of jet power,
    and starts at zero at $t=0$.
    There cannot possibly be a single conversion between radio
    luminosity and jet power, as hoped for in the work of, e.g.,
    \cite{Cavagnolo+10}. The best we could hope for would be a
    relationship between luminosity and power in the mature, large
    phase, which can in fact be seen in simulations
    \citep{English+16}. But even for
    those large sources, the relationship will have significant intrinsic
    scatter dependent on environment.
  \item Because luminosity is dependent on age, environment and, as we
    will see below, particle content, selecting a sample with which to
    calibrate a luminosity-power relationship is fraught with bias --
    a relationship that applies to one group of sources may be
    systematically wrong for another (see Croston et al. 2017
    submitted for evidence that this is in fact the case).
  \item A flux-limited sample, which has a limit on the
luminosity that can be detected at any given redshift, cannot (at
fixed redshift) detect any sources at all below a certain jet power. But, more
importantly, it is also biased against young and old sources. It
should come as no surprise that as we go to more sensitive surveys
we see fewer large, $\ga 100$-kpc scale sources and more smaller,
fainter sources \citep{Baldi+15,Hardcastle+16}.
\end{enumerate}

The aim of the present paper is to use a simple dynamical model of
radio galaxies to make some of the above statements more quantitative. 
Before that, however, I comment on some important observational
results which need to be considered before modelling can be carried
out, and motivate the development of new models by considering
modelling that has been carried out in the past.

\subsection{Key observations: Fanaroff-Riley class, field strength, and particle content}
\label{sec:intro:friii}
The discovery by \cite{Fanaroff+Riley74} that the structures of
powerful sources was directly related to their radio luminosity is of
key importance in our understanding of these objects. In particular,
it demonstrates that the radio luminosity (in these objects, which of
course may not be representative of the whole population) must
correlate at least reasonably well with some more fundamental physical
quantity which also determines the jet structure.

An entirely satisfactory physical explanation of the difference
between the two types of {\it jet} seen in FRI and FRII sources has
existed for many years -- as I will argue in a moment, this is not
quite the same as the difference between the two morphological
classes. The key difference between the jets appears to be their
speed. If we interpret the hotspots that make FRII sources
edge-brightened as terminal shocks, then the jet must be supersonic in
terms of its internal sound speed (for the reverse shock) or the jet
head must be supersonic in terms of the sound speed in the lobes (for
the forward shock). If the lobes are composed of relativistic plasma
(for which see below) then this implies a relativistic bulk speed for
the jet, consistent with directly obtained lower limits on jet speed
from population beaming arguments \citep{Mullin+Hardcastle09}. On the
other hand we know from many very detailed beaming analyses
\citep[summarized by][]{Laing+Bridle14} that the jets in FRI sources
are relativistic on the smallest observable scales but decelerate on
scales of order tens of kpc to sub-relativistic, presumably sub-sonic
speeds. This is entirely consistent with a class of model described by
\cite{Bicknell94,Bicknell96} in which {\it entrainment} plays a
crucial role: all jets necessarily entrain material (from stellar
winds if nothing else: \citealt{Bowman+96,Wykes+15}), and the final
speed of a jet on escaping from the dense central regions in which it
originates is determined by the original jet power $Q$, or more
precisely its momentum flux, and the amount of material it entrains in
its passage through the galaxy. Thus the distinction between FRI and
FRII jets should be related to their jet power $Q$ and the density of
their environment (perhaps by way of the stellar mass of their parent
galaxy; cf \citealt{Ledlow+Owen96}). Since momentum is conserved, it
is almost impossible to evade the necessity for entrainment in order
to produce the observed smooth deceleration of the jets in FRI
sources; stellar mass loss rates and populations are well known and
consistent with producing deceleration from relativistic scales; and,
that being the case, there is arguably no need to consider other
models (e.g.\ of different jet opening angles, magnetic field
structures, or accretion modes) unless a model based on entrainment
alone can be shown to be inadequate. Moreover, there is no real reason
to expect the {\it dynamics} of sources with these two types of jets
to be fundamentally different from the simple model sketched in the
previous subsection\footnote{A very prevalent error is to assume that
  FRI jets, which do not produce terminal shocks inside the jet or
  lobe material, will not drive shocks into the external medium; of
  course these two questions are entirely unrelated, since the
  momentum and energy flux of the jets are unaffected by entrainment.
  Observationally, some of the best examples of shocks driven by radio
  galaxies are in systems classed as FRIs and with
characteristically FRI-like decelerating jets, such as Cen A
  \citep{Kraft+03,Croston+09}.}.

However, it is important here to draw a distinction between the
structure of the jets and the structure of the lobes, a point made
very clearly by \cite{Leahy93}. Let's for convenience label the two
sorts of jet discussed above as `slow' (FRI-type) and `fast'
(FRII-type). We can also classify the extended structures generated by
jets as `lobes' and `plumes', where the difference is essentially one
of aspect ratio; lobes are significantly fatter than plumes, and the
latter often have the appearance of being a smooth continuation of the
jet to large scales. `Classical double' FRIIs of course have fast jets
which feed well-defined lobes, while some of the best-studied FRI
sources, such as 3C\,31, have slow jets that transition seamlessly
into plumes. But a lobed morphology for slow-jet sources is also
extremely common, while somewhat less common, but still easily
identifiable, is the population of `wide-angle tail' sources, which
have mildly relativistic jets \citep{Jetha+08} that often terminate in
clear hotspots \citep{Hardcastle+Sakelliou04} at the bases of long,
extended plumes; these are classically FRI objects from their lack of
edge-brightening but all the evidence is that they possess `fast',
FRII-like jets. Thus both `slow' and `fast' jets can have `lobed' or
`plumed' large-scale structure. The evolution of these two types of
large-scale structure is of course relevant to dynamical models, and
I will return to this point below.

Another key point relating source structures to dynamical issues
concerns the field strength and particle content of radio galaxies.
X-ray inverse-Compton observations of FRIIs
\citep{Hardcastle+02,Kataoka+Stawarz05,Croston+05,Hardcastle+16,Ineson+17}
show conclusively, with very little room for uncertainty, that
magnetic field strength $B$ is typically a small factor below the
equipartition value $B_{\rm eq}$ (where I define equipartition as
with the observed radiating electrons only, i.e. $B^2_{\rm eq}/2\mu_0
= U_e$ where $U_e$ is the energy density in electrons). This departure
from equipartition implies a departure from minimum energy, i.e. the
lobes have a higher energy density, and hence pressure, for a given
synchrotron emissivity than would otherwise be expected. It has also
been clear for some time, but has now been demonstrated in large
samples \citep{Ineson+17} that for these FRII sources, in general, the
pressure from the radiating electron population and field (constrained
by inverse-Compton emission) is very close to the external thermal
pressure from the cluster environment estimated at a radial distance
corresponding to the midpoint of the lobes. This would be a
coincidence if the internal pressure were in fact dominated by
non-radiating particles, and so a plausible view is that non-radiating
particles make a small (though not necessarily negligible)
contribution to the internal pressure in most FRIIs. Such a conclusion
greatly simplifies modelling of such sources since it removes an
ambiguity in the relationship between internal pressure and
synchrotron emissivity.

On the other hand, it is well known that FRI sources generally fall
well below pressure balance at equipartition. This applies not just to
plumed and lobed sources with `slow' jets \citep[e.g.][]{Croston+08}
but also to WAT-type sources with `fast' jets \citep{Hardcastle+05}.
The favoured explanation for this is not a departure from
equipartition \citep{Croston+Hardcastle14} but rather that entrained
thermal material is raised to high temperatures in the lobes and
provides a significant contribution to the internal pressure without
increasing the synchrotron emissivity. This presents a problem for
modelling since the precise properties of this thermal material
(temperature, equation of state) and how these depend on intrinsic
source and environment are not well understood. It is clear, however,
that the synchrotron emissivity of such a source will fall below what
would be expected if the internal pressure were dominated by
electrons. For this reason, among others, population models, which
need to predict radio luminosities, tend to focus on FRII-type
sources, and that is the approach in this paper as well. I emphasise,
however, that the dynamical conclusions of this paper should be
applicable to both types of source.

\subsection{Models of source dynamics and emission}
\label{sec:previous}

\cite{Kaiser+Alexander97} took an important step forward in
constructing analytical models of the evolution of radio galaxies with
what I will refer to as the `self-similar model'; in this model the
aspect ratio of the radio source is constant with time. With this
assumption, and with the further assumption of power-law atmospheres
(i.e.\ number density $n \propto r^{-\beta}$ with $\beta>0$) they were
able to derive analytical expressions for source growth and, later,
synchrotron emission \citep{Kaiser+97}. These models have been
extremely productive of understanding, but they are only as good as
their assumptions. \cite{Hardcastle+Worrall00} argued, based on
observations of radio galaxy environments, that sources may in fact
come close to pressure balance transversely at late times while
continuing to expand longitudinally: in this case the lobe expansion
is no longer self-similar \citep[see also][]{Luo+Sadler10}. Numerical modelling has
allowed us to investigate this scenario quantitatively
\citep{Hardcastle+Krause13,Hardcastle+Krause14} and confirm the
qualitative picture from earlier work; at late times a situation akin
to model C of \cite{Scheuer74} is seen, with the central parts of the
lobe being driven away from the host galaxy by the pressure of the
external medium. In this case, the self-similar assumption breaks
down, which affects source dynamics at late times. At the same time,
the assumption of a power-law environment is not consistent with
observations of clusters, which (unsurprisingly since a power-law
environment is singular at $r=0$) show a flattening of density at
small radii \citep[e.g.][]{Croston+08} affecting the radio
source dynamics at small radii or early times and
invalidating the self-similar assumption by introducing at least one additional
physical scale, the `core radius', to the problem. This second feature
of the self-similar models is particularly problematic since it
prevents them from being applied to modern, realistic density and
pressure profiles of the group and cluster environments that powerful
radio sources inhabit.

Many later versions of analytic models have followed
\cite{Kaiser+Alexander97} in assuming self-similar expansion and/or
power-law environments; this includes the work of \cite{Blundell+99},
\cite{Nath10}, \cite{Mocz+11} and \cite{Godfrey+17}.
\cite{Turner+Shabala15} have developed a formalism which includes
neither of these approximations, at the cost of significantly
increased complexity since they have to solve a large system of
coupled differential equations; their model can, however, be applied to
arbitrary density/pressure profiles. The objective of the present
paper is to develop a model that is conceptually simpler than that of
\cite{Turner+Shabala15} (in the sense that it solves a simpler
  system of equations) without losing the ability to describe the key
physics, encoding the understanding developed from recent numerical
simulations, and to use it with atmosphere models that are well
matched to observations. The model is able to predict the broad-band
integrated synchrotron emission and inverse-Compton emission of both active and
remnant radio galaxies, and in the later sections of the paper I
present some applications to radio galaxy populations.

\section{The model}

\subsection{Model assumptions}

Rather than considering the boundary of the radio-emitting material
explicitly, the model of this paper is based on a simplified
description of the dynamics of the {\it shocked material} around the
radio lobes (the `shocked shell'), from which radio lobe
properties are then inferred. I make the following assumptions:

\begin{enumerate}
  \item Sources have a constant (two-sided) jet power $Q$ throughout
    their lifetime and start at zero size at the centre of a pristine,
    spherically symmetrical environment; thus the radio source is axisymmetric.
  \item The jet is light and relativistic so that its momentum flux in
    one lobe is given by $Q/2c$. (It is easy to modify this assumption
    to deal with slower, heavier jets, and I will make use of this
    fact later.)
  \item The lobes drive a shock into the external medium. The outer
    boundary of the shocked shell takes the form of a prolate spheroid
    with semi-major (`long') axis $R$ and semi-minor axis $R_\perp$.
  \item The unshocked external medium is isothermal (Section \ref{sec:prprof}), i.e. it can be described
    with a single temperature $T$ and therefore constant sound speed
    $c_s$. It has an arbitrary, but spherically symmetrical
    pressure/density profile $p_{\rm ext}(r)$.
  \item {\it A constant fraction} $\xi$ of the energy supplied by the
    jet at any given time is stored as internal energy of the
    relativistic lobe plasma. The remaining energy ($1-\xi$) is stored
    as additional thermal and kinetic energy of the shocked material, between the
    shock front and the contact discontinuity, over and above the
    thermal energy that that material carried across the shock. (This
    assumption is motivated by numerical modelling -- see
    \citealt{Hardcastle+Krause13,English+16} -- where values of $\xi$
    between around $1/2$ and $1/3$ are found.) The lobe material has
    an adiabatic index $\Gamma_j$ and the shocked material has
    $\Gamma_s$: in
    the standard models I take $\Gamma_j = 4/3$ and $\Gamma_s = 5/3$.
  \item The shocked shell is everywhere in (approximate) pressure
    balance with the lobes, as found to be the case in numerical
      models \citep{Hardcastle+Krause14}. The lobes have a single internal pressure
    $p = (\Gamma_j - 1)U = \xi (\Gamma_j-1) Qt/V_L$ (i.e.\ there are no pressure gradients in the lobes).  
  \item The growth of the axes of the shocked shell is governed by the
    non-relativistic Rankine-Hugoniot conditions, in order to
    incorporate mass, momentum and energy conservation at the jump
    between the undisturbed and swept-up gas. In the case of
    transverse expansion, only the internal pressure of the shocked
    gas drives the expansion. In the case of longitudinal expansion,
    the ram pressure of the jet, distributed over the cross-sectional
    area of the lobes, is also relevant.
  \item The lobe long axis (the axis of longitudinal growth as in the
    previous point) is close to the long axis of the shocked
    shell at all times, again as found in numerical models, so that the length of
    one lobe is approximately $R$.
  \item Radiative losses are negligible (in terms of their effects on
    the energetics) at all times -- I test this assumption below.
\end{enumerate}

Assumption (vii) gives us the following differential equations for the
expansion of the shocked shell:
\begin{equation}
\frac{\rd R}{\rd t} = c_s \sqrt{\frac{(\Gamma_s + 1)(p_R/p_{\rm ext}(R))
    - (1-\Gamma_s)}{2\Gamma_s}}
\label{eq:drdt}
\end{equation}
\begin{equation}
\frac{\rd R_\perp}{\rd t} = c_s \sqrt{\frac{(\Gamma_s + 1)(p_{R_\perp}/p_{\rm ext}(R_\perp))
    - (1-\Gamma_s)}{2\Gamma_s}}
\label{eq:drpdt}
\end{equation}
where the pressures to use are given (assumptions (ii) and (v)) by
\begin{equation}
  p_R = \frac{\epsilon QR}{2cV_L} + \frac{(\Gamma_j-1)\xi Qt}{V_L}
\label{eq:pr}
\end{equation}
\begin{equation}
  p_{R_\perp} = \frac{(\Gamma_j-1)\xi Qt}{V_L}
\label{eq:intpr}
\end{equation}
The second of these is the true internal pressure of the lobe, which
by assumption (vi) is the same as the pressure of the shocked
material: the difference between the two pressures arises from
assumption (vii). I estimate the cross-sectional area of the lobe tip
as $V_L/\epsilon R$, by assumption (viii). $\epsilon$ is a geometrical
factor reflecting the fact that the lobe is not cylindrical, so that
the jet momentum flux is not spread over the whole cross-section of
the lobe: I set $\epsilon = 4$ in what follows based on comparison
with simulations. Note here that the internal pressure is almost
always higher than the external pressure, so that the expansion speed
$\rd R/\rd t$ in both directions is higher than the sound speed $c_s$.
In some unusual cases the pressure of the shocked material can fall
below the internal pressure in the perpendicular direction (e.g.\ when
the source is expanding very fast in the longitudinal direction). This
is unphysical, but in this case I force the expansion speed to equal
the sound speed, i.e. the correct speed for the growth of the shell of
gas that can possibly be influenced by the radio galaxy, and the
underpressuring is usually a transient event.

The volume of the lobes $V_L$ is given, as a consequence of assumptions
(v) and (vi), by
\begin{equation}
\frac{V_L}{V_T} = \frac{(\Gamma_j - 1)\xi Qt}{[\xi\Gamma_j +
    (1-\xi)\Gamma_s-1]Qt + NkT - (\Gamma_s -1) \mu N v^2 / 2}
\label{eq:vl}
\end{equation}
where $V_T$ is the total volume inside the shock front (assumption
(iii)):
\begin{equation}
  V_T = \frac{4}{3} \pi R R_\perp^2
\end{equation}
$\mu$ is the mass per particle in the external medium, $v$ is a measure of the
lobe expansion speed (see below), and $N$ is the total number of particles of the external medium that
have been swept up by the expanding shock front, i.e.
\begin{equation}
  N = \int_{V_T} n \rd V
\end{equation}
where the integral is taken over the spheroidal volume and is hence a
function of $R$ and $R_\perp$. (Notice that the lobe occupies a
constant fraction $\xi/(2-\xi)$ of the region behind the shock front
until the swept-up thermal energy becomes comparable to the energy
supplied by the jet.)

These coupled differential equations can be solved numerically for $R$
and $R_\perp$. A numerical integral is of course necessary to compute
$N$ in general, and some initial conditions need to be supplied -- I
take $R, R_\perp = ct_0$ at some small initial time $t_0$. The initial
expansion would be expected to be relativistic but this phase lasts a
comparatively short time, and rather than try to model it accurately
we simply cap the lobe expansion speed at $c$ and use non-relativistic
physics. I have verified that using the more complex expressions
provided by \cite{Gallant} for a strong relativistic shock at early
times makes no significant change to the late-time behaviour. In
addition, we need to make a self-consistent correction for the kinetic
energy of the shocked material, the term in $v^2$ in eq.\ \ref{eq:vl},
since eqs \ref{eq:drdt} and \ref{eq:drpdt} depend on this term through
eqs \ref{eq:pr} and \ref{eq:intpr}. This is implemented,
approximately, by iteratively computing $V_L$ and the expansion speed
and solving using a bisection method for $v^2 = \sqrt{R_\perp (\rd
  R/\rd t)^2+R(\rd R_\perp/\rd t)^2]/(R+R_\perp)}$; the weighted mean
  of the velocities here is intended to take account of the shape of
  the expanding shock front.

This model has some strengths and weaknesses compared to others which
it is worth discussing in detail. Compared to e.g. the analytic models
of \cite{Kaiser+Alexander97}, or the many others discussed above
  that use power-law atmospheres, it has the advantage that it can
deal with any external pressure/density profile, including realistic
ones. It also allows for the evolution of the shock aspect ratio with
time by modelling the transverse and longitudinal expansion
separately. I model both the transition between ram-pressure and
thermal-pressure-dominated expansion (eq.\ \ref{eq:pr}) and also
between the situation where the energy advected across the shock front
is negligible and where it is not (eq.\ \ref{eq:vl}). Unlike the
  model of \cite{Turner+Shabala15}, some of the assumptions break
  down if the expansion speeds become trans-sonic -- since then the
effect of the radio source is not confined to a shocked shell -- and,
in common with all models, it has difficulties if there is
substantial entrainment of thermal material into the lobes themselves,
which would change their equation of state and also affect the
  radio luminosity calculations of the following section. These
  two limitations mean that it is most applicable to powerful sources;
but this is still adequate for the purposes of the remainder of this
paper.

\subsection{Radio luminosity}
\label{sec:radio}

Radio luminosity is computed in the manner applied by
\cite{Hardcastle+Krause13} to numerical models. Suppose we know that the
magnetic field energy density is always some constant fraction of the
electron energy density:
\begin{equation}
  \zeta U_e = U_B = \frac{B^2}{2\mu_0}
\end{equation}
where $\zeta$ describes the energetic departure from equipartition
between field and radiating electrons,
and it is reasonable to assume $\zeta < 1$ (see Section
\ref{sec:intro:friii}). I further assume that the energy density in
non-radiating, relativistic particles (e.g. protons) is given by
$U_{\rm NR} = \kappa U_e$. For a fully tangled field the pressure in the radio
source, $p$, can be related to the energy densities in electrons and
field.
\begin{equation}
p = \frac{U_e + U_B + U_{\rm NR}}{3} = \frac{(1+\zeta+\kappa)U_e}{3}
\end{equation}
or, equivalently,
\begin{equation}
  B = \sqrt{2\mu_0 \frac{3p\zeta}{1+\zeta+\kappa}}
  \label{eq:bfromp}
\end{equation}

Now suppose that the electron energy distribution is a power law in
energy with electron energy index $q$, i.e. 
\begin{equation}
U_e = \int_{E_{\rm min}}^{E_{\rm max}} N_0 E^{1-q} {\rm d}E = N_0 I
\label{eq:pl}
\end{equation}
where
\begin{equation}
I = \left\{ \begin{array}{ll}\ln(E_{\rm max}/E_{\rm min})&q=2\\
{\frac{1}{2-q}} \left[E^{(2-q)}_{\rm max}-E^{(2-q)}_{\rm min}\right]&q\neq
2\\
\end{array}\right .
\end{equation}
$q$ here would be expected to be set by particle acceleration
processes: for strong shocks we might expect $q\sim 2$, but the
particle acceleration physics is not part of the model, so $q$ is a
parameter that should be set by the user. $I$ is only weakly dependent
on $E_{\rm min}$ and $E_{\rm max}$ for plausible values of $q$. I use
values corresponding to $\gamma_{\rm min} = 10$, $\gamma_{\rm max} =
10^6$ in what follows. The choice of $\gamma_{\rm min} = 10$,
  which has been used in much previous work, is driven
by the suggestion of low-energy cutoffs around Lorentz factors of a
few hundred in hotspots \citep[see e.g.][for a compilation]{Hardcastle04}
together with some assumptions about adiabatic expansion from the
scale of the hotspots and the lobes,
although it now seems plausible that the detailed hotspot energy
spectrum is not consistent with a sharp cutoff \citep{McKean+16}. The
difference over the range $1 < \gamma_{\rm min} < 100$ is negligible
compared to other uncertainties for $q\approx 2$.

It can be shown \citep[e.g.][]{Longair10} that for a power-law distribution
of electrons, the volume spectral emissivity of synchrotron radiation is given
by
\begin{equation}
J(\nu)  = C N_0 \nu^{-\frac{(q-1)}{2}} B^{\frac{(q+1)}{2}}
\label{jnu}
\end{equation}
where
\begin{equation}
C = c(q) {\frac{e^3}{\epsilon_0 c m_e}} \left({\frac{m_e^3 c^4}{e}}\right)^{-
(q-1)/2}
\end{equation}
($c(q)$ here is a dimensionless constant, of order 0.05 for plausible
values of $q$ -- see \citealt{Longair10} for details.)

Therefore (since $N_0 = U_e/I$) the total radio luminosity per unit frequency can be
written
\begin{equation}
  L_{\rm radio} = J(\nu)V = \frac{C}{I} \frac{E_{\rm
      lobe}}{1+\zeta+\kappa} \nu^{-\frac{(q-1)}{2}}
  B^{\frac{(q+1)}{2}}
  \label{eq:lr}
\end{equation}
where $B$ is given by eq.\ \ref{eq:bfromp} applied to the internal
pressure in the lobe (eq.\ \ref{eq:intpr}), and $E_{\rm lobe} = \xi Qt$.

One can see that a substantial energy density in non-radiating particles
($\kappa \ga 1$) reduces both $B$ (eq.\ \ref{eq:bfromp}) and the
normalization of $L_{\rm radio}$. As noted above, for FRI sources there is evidence
that non-radiating particles entrained from the environment in the
process of jet deceleration have a significant role,
and in that case the radio luminosity will be lower for a given jet
power and lobe volume. (Although on the face of it eq.\ \ref{eq:lr} is
independent of lobe volume, a volume dependence enters through the
dependence on $B$, which depends on energy density.) The standard
assumption in the rest of the paper will be that, for the FRII sources
I aim to model, $\kappa = 0$.

\subsection{Loss processes}
\label{sec:loss}

It is very important to take account of the effects of radiative and
adiabatic losses on the observed synchrotron radiation. Since we are
assuming that $Q$ is constant (assumption (i) above) we can make the
related (though not identical) assumption that the supply of
high-energy particles in power-law form is constant with time, as done
by \cite{Kaiser+97} and papers following that formalism. Then
the true electron energy spectrum of the lobes at time $t$ is the
integral
\begin{equation}
  N(E) = \int_0^t N_{\rm aged}(E,t_{\rm inj},B_{\rm eff}) \rd t_{\rm
    inj}
  \label{eq:aged}
\end{equation}
where $N_{\rm aged}$ is the suitably normalized electron energy
spectrum of a population of electrons injected at time $t$ and $B_{\rm
  eff}$ is the effective magnetic field strength that has aged those
electrons, taking into account the time variation of $B$ and losses to
inverse-Compton scattering of the cosmic microwave background (CMB) --
that is, $B_{\rm eff}$ is itself a time average of $B^2$ and
$B^2_{CMB}$ between $t_{\rm inj}$ and $t$. The synchrotron spectrum of that
population can then be calculated using the current value of $B$ and
used to derive a correction for the pure power-law spectrum of
eq.\ \ref{eq:pl}. For these purposes I use simple
\cite{Jaffe+Perola73} aged spectra\footnote{As discussed by
  \cite{Hardcastle13}, alternatives that do not involve pitch angle
  scattering of the electrons, such as the widely used
  Kardashev-Pacholczyk models \citep{Kardashev62,Pacholczyk70} are not
  realistic; we would obtain similar results by using the more
  physically realistic \cite{Tribble91} models but at significantly
  increased computational cost.} and compute the synchrotron spectrum
of aged populations at a finite set of discrete times in order to
approximate the emission from the electron population of
eq.\ \ref{eq:aged}, using a version of the code of
\cite{Hardcastle+98-2}. Note that I am assuming here that each element
of the electron population contributes equally to the synchrotron
emissivity: this may not be the case if, e.g., there is magnetic field
structure in the lobe.

Adiabatic losses may be represented as a further correction to the
radiative loss factor. Adiabatic expansion from volume $V_0$ to volume
$V_1$ reduces the energies of all electrons in the volume: $E \propto
(V_1/V_0)^{-1/3}$. As the characteristic electron energy for aged
electrons goes as $t^{-1}$, we can correct approximately for adiabatic
effects on the spectrum by increasing the effective age of a
population by a factor of the cube root of the ratio of the volume of
the lobes at injection to the currently observed volume, as if the
particle population had first aged and then abruptly expanded to their
current volume. This is only an approximation, since in reality the
lobe expansion and the radiative losses occur in parallel, but the
correction turns out to be a small one in any case. The effects of
adiabatic losses on the energy density in the lobes are assumed to be
taken account of by assumption (v).

\subsection{Inverse-Compton emission}

Inverse-Compton emission due to scattering of the CMB, the other main
detected loss process of radio galaxies, can be modelled in a very
similar way to synchrotron emission; to determine the inverse-Compton
emissivity I integrate over the full electron and photon distribution
with a suitable kernel in the manner described by \cite{Hardcastle+98-2}. With a simple power-law
assumption for the electron energy spectrum, this process gives an inverse-Compton luminosity that
increases linearly with time for $t<T$ and is independent of environment, because volume emissivity is
proportional to the normalization of the electron spectrum, and
\begin{equation}
  N_0 V \propto \frac{\xi Qt}{{1+\zeta+\kappa}}
\end{equation}
With corrections for loss and adiabatic expansion as described in the
previous subsection, it is possible for electrons of the required
energy for inverse-Compton scattering to a particular energy (e.g., $\gamma
\sim 10^3$ to scatter the CMB into the soft X-ray at $z=0$) to be removed, introducing an
environment dependence of the inverse-Compton luminosity.

I do not consider nuclear inverse-Compton or synchrotron self-Compton
processes in this paper either as a radiative loss term or as a source
of observable photons. The former would add a dependence on the accretion state of
the AGN generating the jet, which complicates the picture
significantly, and is in any case only important for small sources
because of the $1/r^2$ dependence of the photon energy density. The
latter is never dominant over synchroton emission.

\section{Model results}

\subsection{Realistic model atmospheres}
\label{sec:prprof}
I select two model atmospheres in the current paper for the purposes of testing.

The first is the widely used isothermal beta model
\citep{Cavaliere+Fusco-Femiano78}, where we have
\begin{equation}
  p = p_0\left[
      1+\left(\frac{r}{r_c}\right)^2\right]^{-3\beta/2}
\end{equation}
and $n=p/kT$. The smoothness and simplicity of this model makes it a
quick and easy basis for testing and results that use it can also be
directly compared with the numerical models of HK13, HK14. However, it
is not particularly well motivated either by theory or observation,
and has four free parameters ($p_0$, $kT$, $r_c$, $\beta$) which do
not map in a particularly obvious way on to the richness of a group or
cluster (other than in the obvious qualitative sense that larger
$p_0$, $kT$ or $r_c$ imply more gas).

The second is the so-called universal pressure profile of
\cite{Arnaud+10}. This is derived from the observed pressure profiles
of a well-studied sample of clusters, and has the great advantage that
it is calibrated in terms of a single free parameter, the total mass
of the system, $M_{500}$ (the total gravitating mass within the radius
$R_{500}$ corresponding to a density contrast of 500 times the
critical density of the Universe) -- a given $M_{500}$ uniquely specifies the pressure
profile. Although this relationship was calibrated for clusters with
$10^{14} < M/M_\odot < 10^{15}$, \cite{Sun+11} have shown that it can
be extended to groups, and therefore it applies across the range of
environments known for powerful radio galaxies. I implement the
universal pressure profile using the prescription in section 5 of
\cite{Arnaud+10}, and then assume an isothermal temperature profile
with $kT$ from the $M_{500}$-$kT$ relation of \cite{Arnaud+05} in
order to convert to density for model testing (incorporating a
temperature profile would be trivial in the context of the modelling
but is not necessary at this point).

\begin{figure}
  \includegraphics[width=1.0\linewidth]{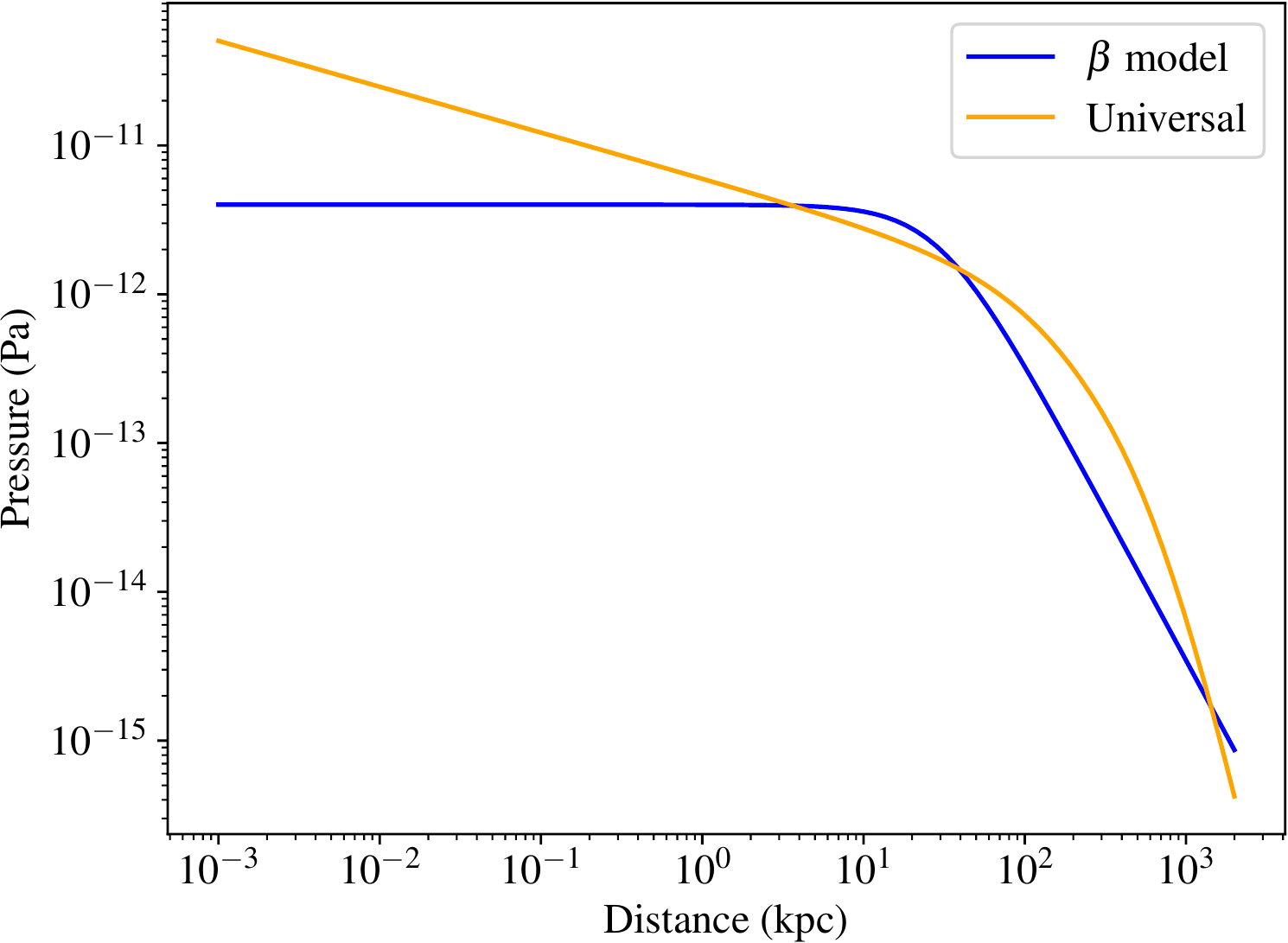}
\caption{Pressure profiles for a representative $\beta$ model and
  \protect\cite{Arnaud+10} universal pressure profile, as described in
the text.}
\label{fig:prprof}
\end{figure}

Fig.\ \ref{fig:prprof} shows a comparison between a universal pressure
profile with $M_{500} = 10^{14} M_\odot$ and a $\beta$ model with
$p_0=4 \times 10^{-12}$ Pa, $r_c=30$ kpc, $\beta = 2/3$ and $kT =
2.3$ keV, both intended to represent a poor cluster environment not
uncommon for radio galaxies \citep{Ineson+15}. It can be seen that the
most important difference is in the pressure (therefore density) at
small radii, $<1$ kpc, but there are non-negligible differences at
large radii too.

\subsection{Simple source dynamics}
\label{sec:dynamics}

\begin{figure*}
  \includegraphics[width=1.0\linewidth]{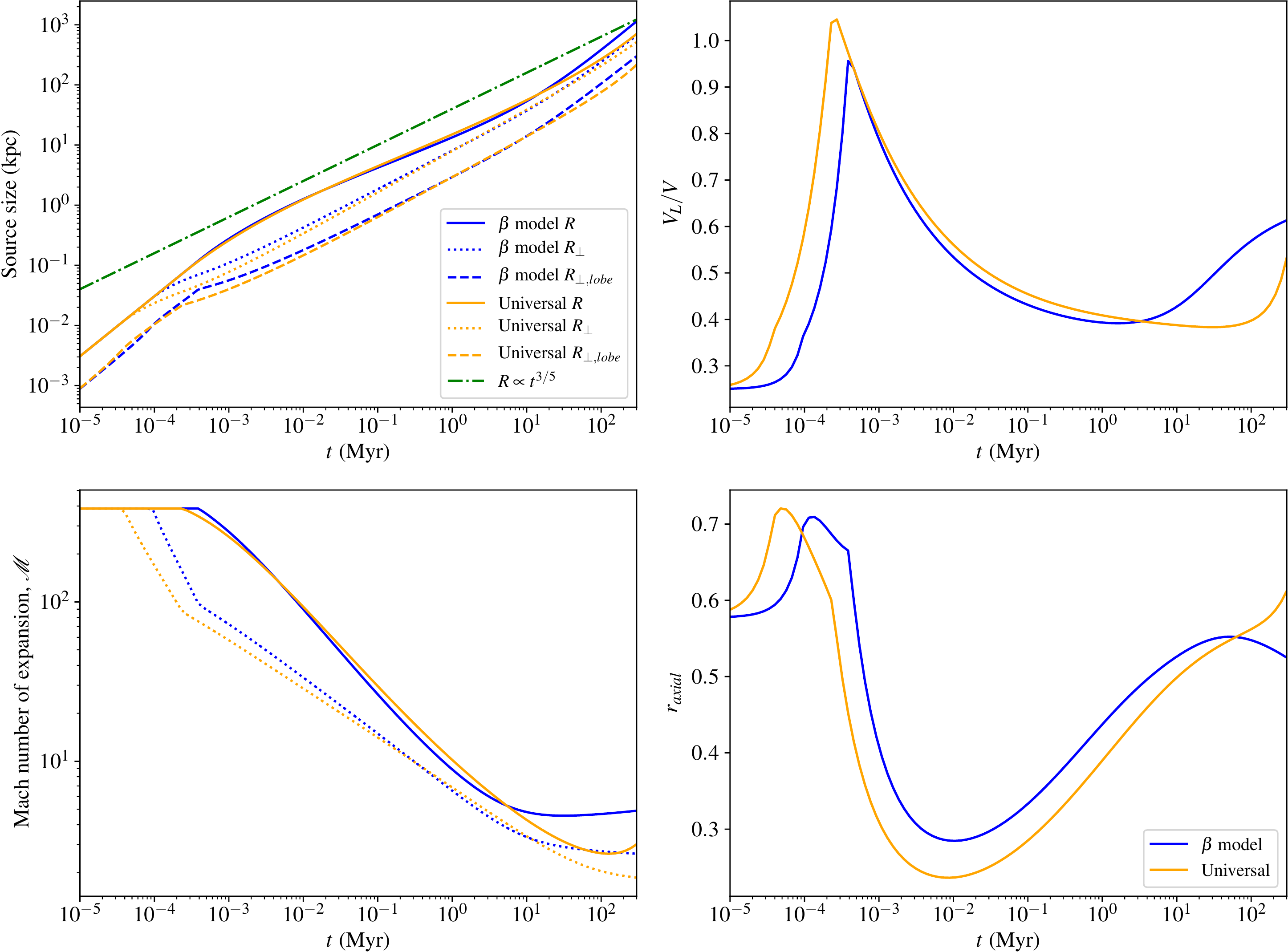}
\caption{Dynamical quantities for a jet of $Q = 2\times
10^{39}$ W in two poor cluster environments as described in the text.
Top left: source length $R$, perpendicular source radius $R_\perp$,
and perpendicular lobe radius $R_{\perp,lobe}$: an arbitrarily
normalized line of $R \propto t^{3/5}$ is also shown. Top right: lobe volume
as a function of total volume, $V_L/V$. Bottom left: Mach number of
the longitudinal and perpendicular advance of the shocked region.
Bottom right: the axial ratio of the radio source
$2R_{\perp,lobe}/R$.}
\label{fig:simple}
\end{figure*}

In this section I use the two model atmospheres from Section
\ref{sec:prprof} to present some basic results of the modelling. I
model the propagation of a light relativistic jet with $Q = 2\times
10^{39}$ W into both atmospheres and solve for times in the range 0 --
300 Myr. The power is chosen to match the intermediate power used in
the numerical simulations of \cite{English+16}, and I set $\xi =
0.4$, also matching their results. Fig.\ \ref{fig:simple} shows some
key dynamical quantities from the results.

The top-left plot of Fig.\ \ref{fig:simple} shows the source expansion
with time: the basic features of this plot are common to all modelled
sources. Expansion is originally relativistic but transitions to
sub-relativistic lobe advance speeds on short timescales, $\sim 1000$
years in this case. It is important to remember that the model is not
accurate when the expansion speed is relativistic. There then follows
a phase of decelerating expansion (compare the bottom-left plot, which
shows the Mach number of expansion): in the $\beta$-model environment,
where the density profile is essentially flat on these scales, the lobe growth has the
expected $R \propto t^{3/5}$ slope \citep{Kaiser+Alexander97}. The
lobe advance speed then changes (flattening off in the $\beta$-model
case or accelerating in the universal pressure profile case) when the
front of the lobe starts probing the steep pressure gradient on scales
of $\sim 100$ kpc. It is also worth noting the
comparatively slow growth of these model sources -- in 300 Myr the
source has grown only to Mpc scales, implying an average growth speed
of $\sim 0.01c$. Although the figure clearly depends on the
environment and jet power, slow jet head advance speeds are a feature
of light jets such as those used here.

The expansion is clearly not self-similar -- note
both the modest changes in the fraction of the volume occupied by the
lobe in the top-right plot and the much larger changes in the source
axial ratio in the bottom right. The axial ratio behaviour we see here can
qualitatively be understood in terms of the different density regimes
and the evolution of pressure in the lobes, which lead to different
expansion speeds at different times. At very early times, the lobes
are expanding relativistically in all directions, and so we see almost
spherical lobes with large axial ratios. Once this phase is over, the
lobes grow linearly much faster than they do transversely because the
ram pressure term in eq.\ \ref{eq:pr} dominates over internal
pressure: however, as the source expands the ram pressure term becomes
less important, and so at later times the transverse and longitudinal
expansion speeds become more similar and consequently (integrated over
time) the lobe linear and transverse sizes become more similar. At the
very latest times shown in the figure, the tips of the lobe start to
probe the very steep downwards density/pressure gradient in the
outskirts of the cluster and consequently the longitudinal expansion
starts to accelerate while the transverse expansion continues to
decelerate: this leads to a decrease in the axial ratio again.

\subsection{Comparison with simulations}

\begin{figure*}
  \includegraphics[width=1.0\linewidth]{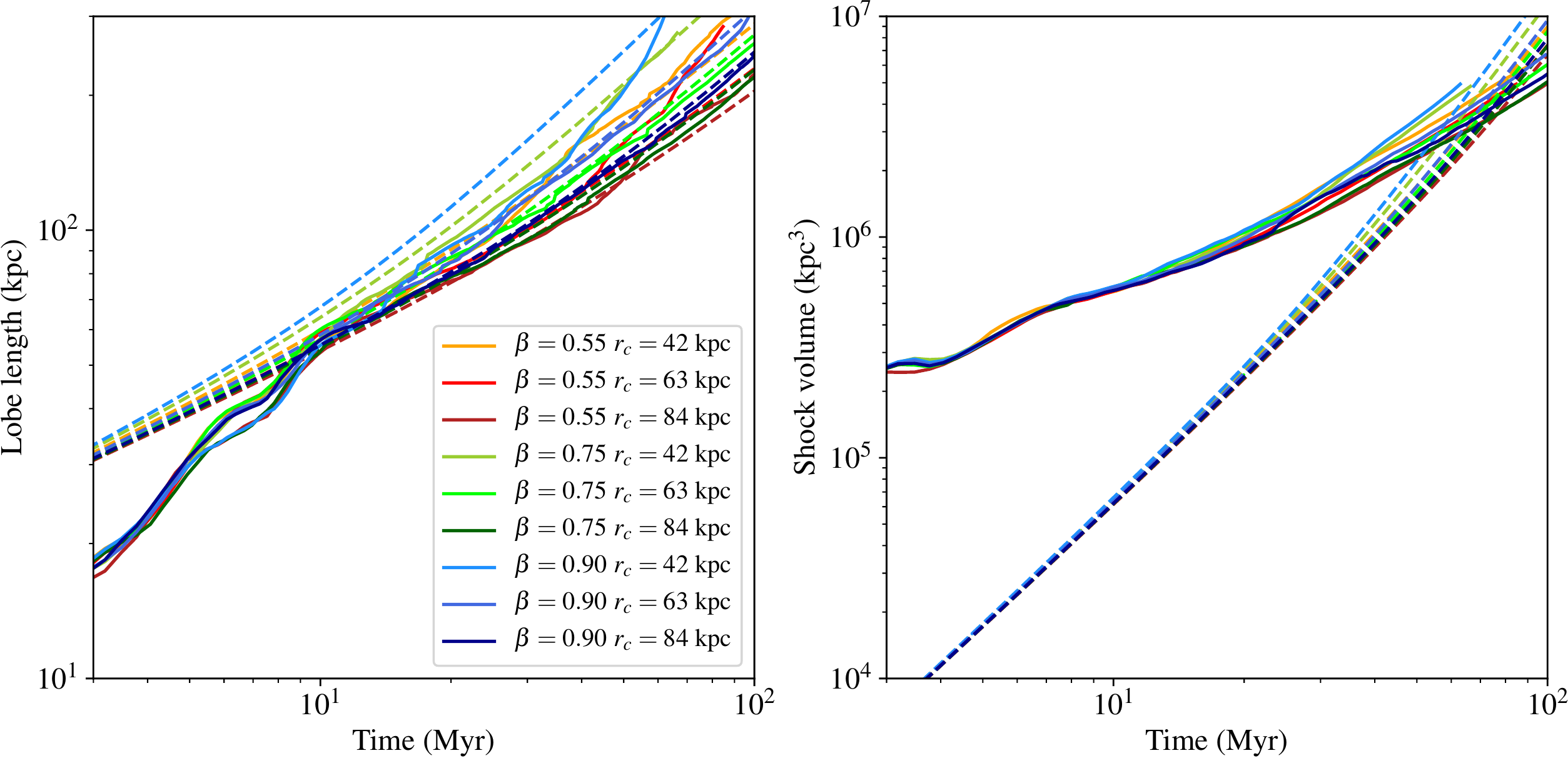}
\caption{Analytic models compared with the HK13 2D heavy jet numerical
  models. Left: lobe/shocked region length. Right: shocked region
  volume. Solid lines show the hydrodynamical simulations, dotted
  lines show the analytical models.}
\label{fig:hk13-compare}
\end{figure*}

For a direct comparison with simulations I model jets in the set of
environments used in the numerical models of HK13, i.e. $\beta$ models
with $kT=2$ keV, $p_0=10^{-11}$ Pa, $\beta = [0.55, 0.75,0.90]$ and
$r_c = [20,30,40] \times 2.1$ kpc. I omit the unrealistically flat
$\beta = 0.35$ models for simplicity and use a two-sided jet power of
$2 \times 10^{38}$ W. In order to match HK13's jet conditions I
adjust the jet momentum flux so that it is appropriate for a ${\cal M}
= 25$ jet of this power (a `heavy jet'), and set the adiabatic index of the jet
material $\Gamma_j$ to 5/3.

Fig.\ \ref{fig:hk13-compare} shows comparisons between (left panel)
the lobe lengths and (right
panel) the volume of the shocked region, in simulations and analytic
models. It can be seen that the analytical models agree well with the
simulations at late times on the expected lobe lengths: at early times
the simulations are affected by the pixel size in the simulation and
by problems coupling the boundary-condition jet to the ambient medium.
The analytical models show the same late-time qualitative trends and
quantitatively agree to within $\sim 20$ per cent for most
simulations. It is important to note here that for these heavy jets
the jet momentum flux is still important in driving the forward
expansion of the lobes at late times, so the agreement here is not
particularly surprising, particularly as the parameter $\epsilon$ has
been chosen to produce general agreement with simulation results.

The shock volumes in the analytical and numerical models also agree
reasonably well at late times. The numerical model volumes are
unrealistically high in volume at early times (due to resolution
  effects and to the use of a boundary-condition jet, the shock in the
  numerical models effectively has non-zero size at $t=0$), but at
later times they are close to consistent, though with a different
gradient. The numerical models do predict smaller shocks at the very
latest times. Part of this may be due to the effect of the lobes
pinching off in their central regions and ceasing to drive transverse
expansion, an important feature of the numerical models which the
analytical models do not include. This coincides with a transition to
trans-sonic or sub-sonic expansion in the analytical models, and I
have already noted that this regime is unlikely to be modelled well.
As a consequence, the lobes themselves are larger and (since by
construction the total energies in the two regions have a constant
ratio) the pressure in the lobes and shocked regions is smaller than
in the numerical models at a given late time.

\subsection{Synchrotron emission and losses}
\label{sec:results_loss}

\begin{figure*}
  \includegraphics[width=1.0\linewidth]{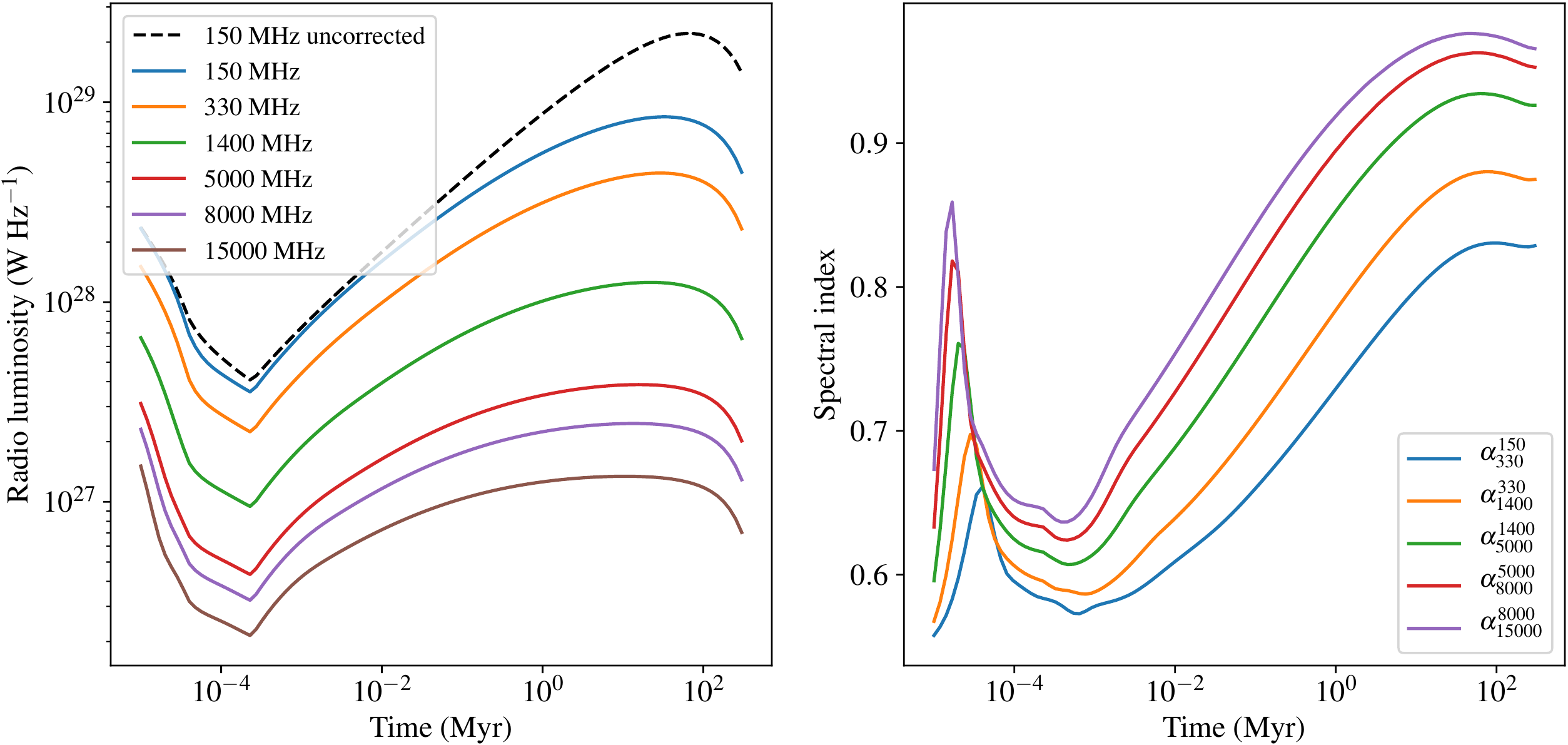}
  \includegraphics[width=1.0\linewidth]{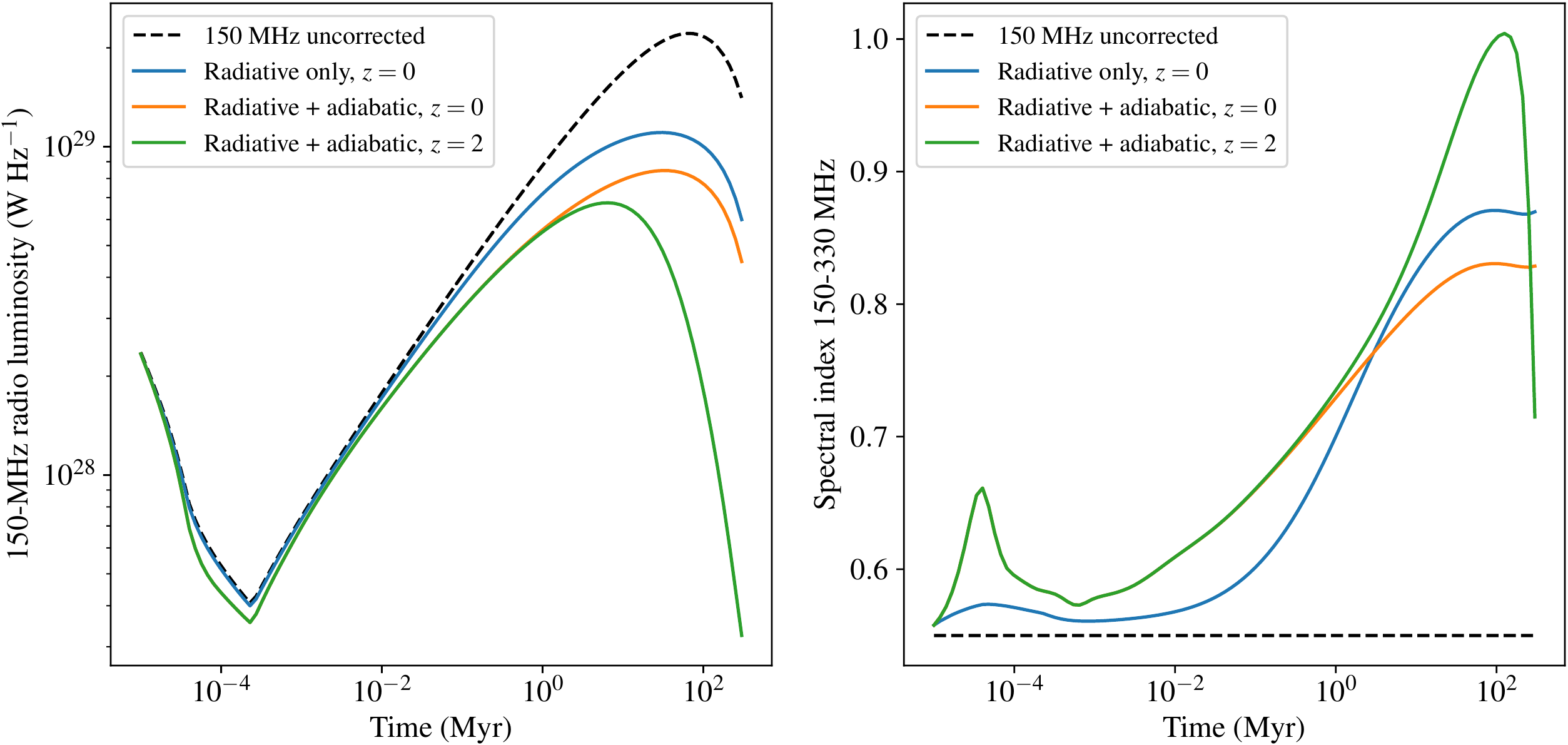}
\caption{Synchrotron emission as a function of time for a universal
  pressure profile with $M_{500} = 10^{14} M_\odot$ and a light jet
  with $Q = 2\times 10^{39}$ W. Top two panels: coloured lines show (left) radio
  luminosity at different frequencies and (right) spectral index
  between adjacent pairs of frequencies at $z=0$ and with
  corrections for spectral ageing and adiabatic losses, with the uncorrected 150-MHz
  emission luminosity plotted for reference. Bottom two panels: the
  radio luminosity at 150 MHz, and the spectral index between 150 and
  330 MHz, on different assumptions about the redshift and the
  inclusion of the adiabatic expansion correction.}
\label{fig:synchrotron}
\end{figure*}

The top panel of Fig.\ \ref{fig:synchrotron} shows the synchrotron
emission as a function of time expected for the universal pressure
profile, $Q=2 \times 10^{39}$ W system described in Section
\ref{sec:dynamics} for various frequencies commonly used in radio
astronomy. Also plotted is the integrated spectral index $\alpha$
(defined in the sense that $S \propto \nu^{-\alpha}$) for each
adjacent pair of spectra. I set $q=2.1$, corresponding to an
`injection index' $\alpha = 0.55$ (which would be expected in
  models of particle acceleration at strong shocks), and assume $z=0$ and $\zeta=0.1$,
the latter based on the results of \cite{Croston+05} --
unlike \cite{Blundell+99} I do not try to model any dependence of the
injection index on jet power. I also plot the 150-MHz luminosity
uncorrected for any spectral age or adiabatic loss effects, i.e.
derived purely as discussed in Section \ref{sec:radio}.

This plot has several important features. Firstly, we can see that as
seen in numerical modelling (HK13, HK14, E16), but unlike what is the
case in analytical models based on the assumption of a power-law
atmosphere \citep{Kaiser+97,Mocz+11} the synchrotron luminosity starts
low, rises to a peak, and then falls off again. (The peak in
  luminosity at very early times, $t \la 10$ yr, comes from the phase
  of the source's life in which its expansion speed is limited to the
  speed of light: given the approximations used in this regime it should
  probably not be taken too seriously.) The evolution of the radio
luminosity seen in these plots between times of $\sim 10^{-2}$ and
$10^2$ Myr is generally to be expected from any model with a
pressure/density profile which starts relatively flat and then
steepens, together with an equipartition-type assumption. (Note that
the luminosity at very early times would be affected by synchrotron
self-absorption, which is not modelled here.) The fall in synchrotron
luminosity at late times (which starts to become obvious when the
source size approaches the linear size at which the pressure gradient
slope changes) is made more prominent, but {\it not} caused, by the
effects of radiative and adiabatic losses, as can be seen from the
uncorrected light curve.

Secondly, we see different evolution of the light curves at different
frequencies. The integrated spectrum of the source is expected to be
curved downwards at all frequencies (here, unlike in real sources,
there are no confusing effects of beamed flat-spectrum jets or
self-absorbed cores) but the curvature changes with time. The
low-frequency integrated spectral index of the source is steeper than
the injection index at effectively all times (again, it is important
to note that the intrinsic spectral index at very early times would be
rendered unobservable by self-absorption). Note also that,
self-absorption effects aside, the low-frequency light curve is the
one that changes {\it most} with time -- as expected since it is
affected most strongly by the evolutionary history of the source,
which tends to be erased by radiative losses at high frequencies. In
reality, of course, high-frequency light curves will be affected by
beaming and by transient features such as hotspots which are not
modelled here. Another important feature of the models is that
spectral index does not steepen monotonically with time -- at late
times, after the onset of rapid expansion of the source, we actually
see a flattening of the spectral index (which is likely due to the
fact that at late times very old material has an increasingly
negligible effect on the integrated spectrum). At no point in the
source evolution is the spectrum over this range modelled well as a
broken power law or `continuous injection (CI)' model
\citep{Kardashev62}, consistent with some of the sources
  discussed by \cite{Harwood17}. The deviation from the CI expectation
  comes about because the loss rates (both to adiabatic and radiative
  losses) are not constant with time in the source modelled here, and
  will be reduced for large, old sources where inverse-Compton
  radiative losses dominate, which may help to explain why
  \cite{Harwood17} found some sources to be reasonably well described
  by the CI model. Of course, it is important to bear in
mind that the detailed spectral index behaviour depends crucially on
the assumption that the magnetic field energy density is a constant
fraction of the total, something that we do not know to be true for
real radio sources, and on the assumed value of the injection
  index $q$ and the energy density to electron energy ratio parameter
  $\zeta$.

The normalization of the radio luminosity that is found here is very
consistent with the numerical results of E16 for light jets of the
same power, which do not take into account radiative losses but
  which of course make similar observationally based assumptions about
  the injection index and magnetic field to electron energy density
  ratios. More importantly, it is consistent with the
observed jet power/radio luminosity relation given by
\cite{Ineson+17} (bearing in mind that the sources in the latter
paper, by selection, tend to be large, powerful objects which will be
close to the peak of their light curves: I return to this point below,
Section \ref{sec:ensemble-jetpower}).

To make the effects of the different assumptions clear, the bottom
panel of Fig.\ \ref{fig:synchrotron} shows the 150-MHz luminosity, and
150-330 MHz (rest-frame) spectral index, for four different
combinations of assumptions: no corrections; $z=0$ spectral ageing
only; $z=0$ spectral ageing and adiabatic losses (as in the top panel)
and $z=2$ spectral ageing and adiabatic losses. The comparison of the
first three of these demonstrates, as noted in Section \ref{sec:loss},
that the effect of the approximate correction that I make for
adiabatic losses is relatively small, though not negligible, on both
the total luminosity and integrated spectral index; the effect of
spectral ageing alone is more significant. By contrast, comparing the
$z=0$ and $z=2$ curves, we can see that the effects of going to high
redshift (so increasing losses to the CMB) are very significant, as
noted by \cite{Kaiser+97}. These losses start to become important when
the energy densities in magnetic field and CMB photons are comparable,
which for the modelled source occurs after only a few Myr at $z=2$. A
very striking effect is seen in the spectral index plot, where we see
that the high-$z$ source does indeed have a steeper spectrum than at
low $z$ over some of its evolution, but at later times has a flatter
spectrum. This is because inverse-Compton losses have removed almost
all of the aged electrons, leaving only flat-spectrum material that
has recently been injected. The well-known association between higher
$z$ and steeper spectrum (if it exists at all and is not just a
luminosity-$\alpha$ relation) might therefore be a selection effect in
flux-limited samples.

\subsection{Remnant sources}
\label{sec:remnant}

It is clear that the model above can relatively easily be modified to
deal with non-constant jet power $Q$, e.g. by replacing terms in $Qt$
by an integral of $Q$ over time. In general we know too little about
variations in $Q$ with time in real sources to make this a worthwhile
exercise. However, remnant sources, where the jet turns off in the course of the source's
evolution, are an important special case which is easy to implement.
Here effectively we need to consider
\begin{equation}
  Q = \left\{ \begin{array}{ll}Q_0&0<t<T\\
    0&t\ge T\\
  \end{array}\right .
\end{equation}
The ram pressure of the jet also drops to zero at $t>T$. If we retain
all the other assumptions, then the dynamics alter: we have
  \begin{equation}
  p_R =
  p_{R\perp} = \xi (\Gamma_j - 1) QT/V_L
  \end{equation}
for $t>T$. Similar changes need to be made to deal with the term for
the internal energy of the lobes $\xi Qt$ wherever else it appears,
e.g. in the computation of radio luminosity. Finally, the radiative
loss code needs to be modified to take account of the fact that
particle acceleration will cease, i.e. no new particles will be
injected for $t>T$. Although it seems likely that in reality others of
the model assumptions will break down (e.g. the constancy of $\xi$ as the
source expands) these modifications are sufficient to give an
impression of the expected evolution of a remnant source.

\begin{figure*}
  \includegraphics[width=1.0\linewidth]{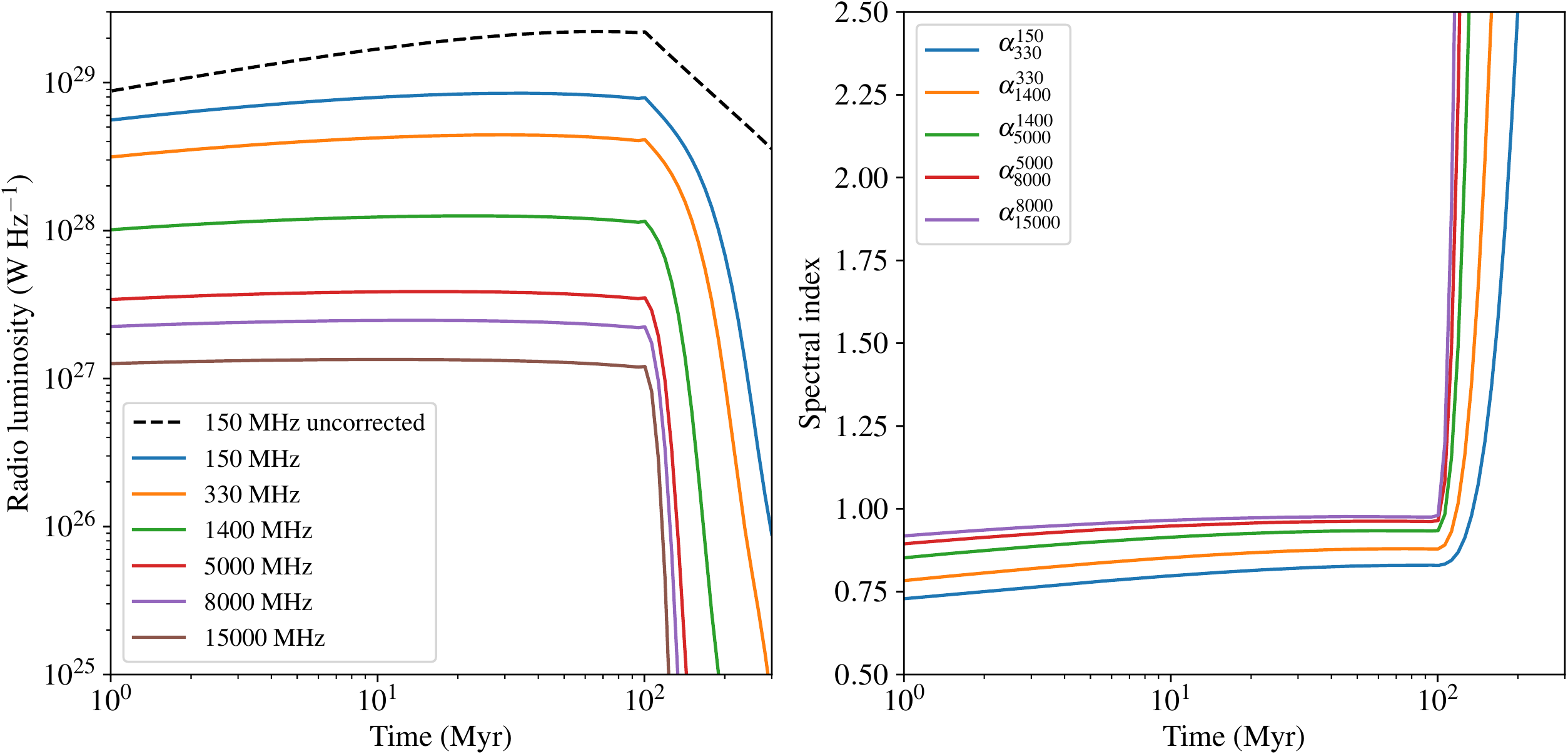}
\caption{Synchrotron emission as a function of time for a universal
  pressure profile with $M_{500} = 10^{14} M_\odot$ and a light jet
  with $Q = 2\times 10^{39}$ W at $z=0$, switched off after 100 Myr and allowed
to evolve for a further 200 Myr. Lines as in
top two panels of Fig.\ \ref{fig:synchrotron}, but note the different scales on the axes.}
\label{fig:synch-remnant}
\end{figure*}

To examine remnant evolution I take the same case of a powerful ($Q =
2 \times 10^{39}$ W) jet in a uniform pressure profile with $M =
10^{14} M_\odot$ at $z=0$, but now allow the jet to switch off after
100 Myr. The results are shown in Fig.\ \ref{fig:synch-remnant}, which
can be compared directly to Fig.\ \ref{fig:synchrotron}, although I
now focus on the time evolution after 1 Myr. One can see a very rapid drop
in the synchrotron luminosity of the source at all frequencies
immediately after the jet is disconnected, as also seen in the quite
different models of \cite{Godfrey+17}. Two effects contribute to this
drop. Firstly, the dynamics of the source change, as the drop in lobe
pressure due to the continued expansion at the lobe tip is no longer
counteracted by the energy supply: this leads to a drop in $B$ as a
function of time. Secondly, and dominantly as can be seen by comparing
the corrected and uncorrected light curves, radiative and adiabatic
losses are no longer offset by a continued injection of young
particles. The effect is an almost instantaneous disappearance of the
high-frequency emission. Even low-frequency emission drops by an order
of magnitude over 100 Myr. Loss effects would of course be much more
significant at higher redshift where inverse-Compton losses dominate
(see Fig.\ \ref{fig:ic-remnant} for an example of this).
The timescales for a significant drop in luminosity will depend on the
jet power and environment, but it seems clear that this rapid fading
of the lobes can contribute to the small fraction of remnant radio
galaxies seen at low frequencies (for recent constraints on this
fraction see \citealt{Brienza+17}, \citealt{Godfrey+17} and Mahatma et
al.\ (submitted)). I discuss this point further below, Section
\ref{sec:ensemble-remnant}.

\subsection{Inverse-Compton light curves}

\begin{figure*}
  \includegraphics[width=1.0\linewidth]{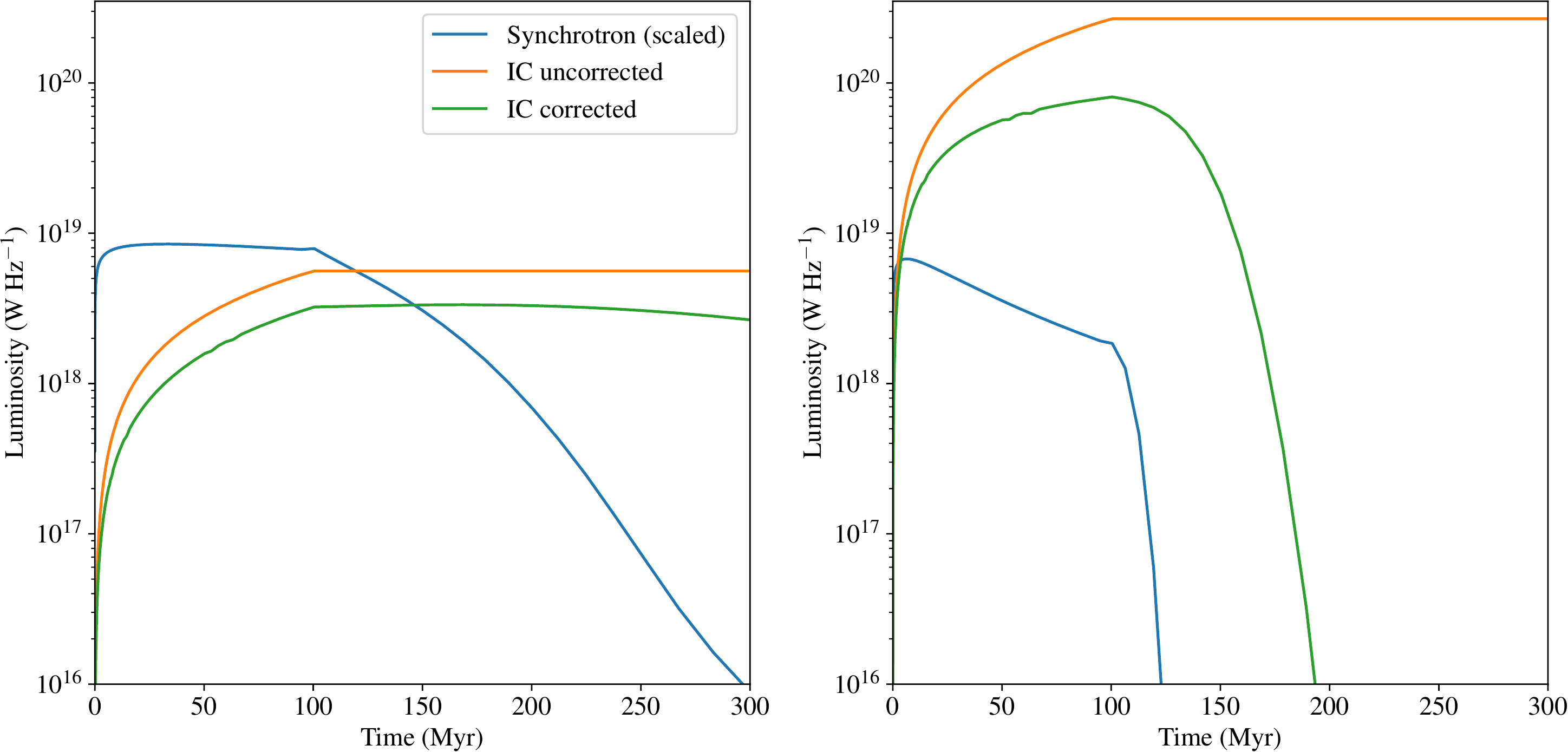}
\caption{Inverse-Compton and synchrotron emission as a function of
  time for the remnant source of Fig.\ \ref{fig:synch-remnant}. Here I
  use a linear time scale to emphasise the decay phase. Synchrotron
  emission is calculated at {\it rest-frame} 150 MHz and inverse-Compton at
  rest-frame 1 keV ($2.4 \times 10^{17}$ Hz). The synchrotron
  luminosity has been scaled down by a factor $10^{10}$. Left: $z=0$.
  Right: $z=2$.}
\label{fig:ic-remnant}
\end{figure*}

As noted above, uncorrected inverse-Compton light curves are linear
with time until the jet turns off. Correction factors for expansion
and radiative losses are expected to be non-negligible, though,
because electrons injected at early times will only contribute at very
low energies. After the jet switches off, the inverse-Compton emission
will decline. Fig.\ \ref{fig:ic-remnant} shows example inverse-Compton
light curves for the remnant source of the previous section, with loss
and adiabatic corrections applied, for $z=0$ and $z=2$. In both cases
we see that the decline of inverse-Compton emission with time is much
slower than the synchrotron light curve, as expected since the
synchrotron emission is due to higher-energy electrons. Both
qualitatively and quantitatively these curves are very similar to
those found by \cite{Mocz+11} -- compare in particular their figure 3
with the right-hand panel of Fig.\ \ref{fig:ic-remnant} -- apart from
the unrealistically high synchrotron emission at early times in their
figure, a result of their assumptions about environment. This supports
their prediction of inverse-Compton `ghosts' associated with the
remnants of powerful sources, particularly at low $z$ where the
decline in inverse-Compton emission is very slow. One might expect to
find these associated with steep-spectrum radio remnants which could
be identified in e.g. LOFAR surveys. Of course the X-ray surface
brightness of these structures will be low at low redshift, where
their angular size is large, and distinguishing their emission from
thermal emission from the environment may prove challenging.

\subsection{Integrated losses and self-consistency}

\begin{figure}
  \includegraphics[width=1.0\linewidth]{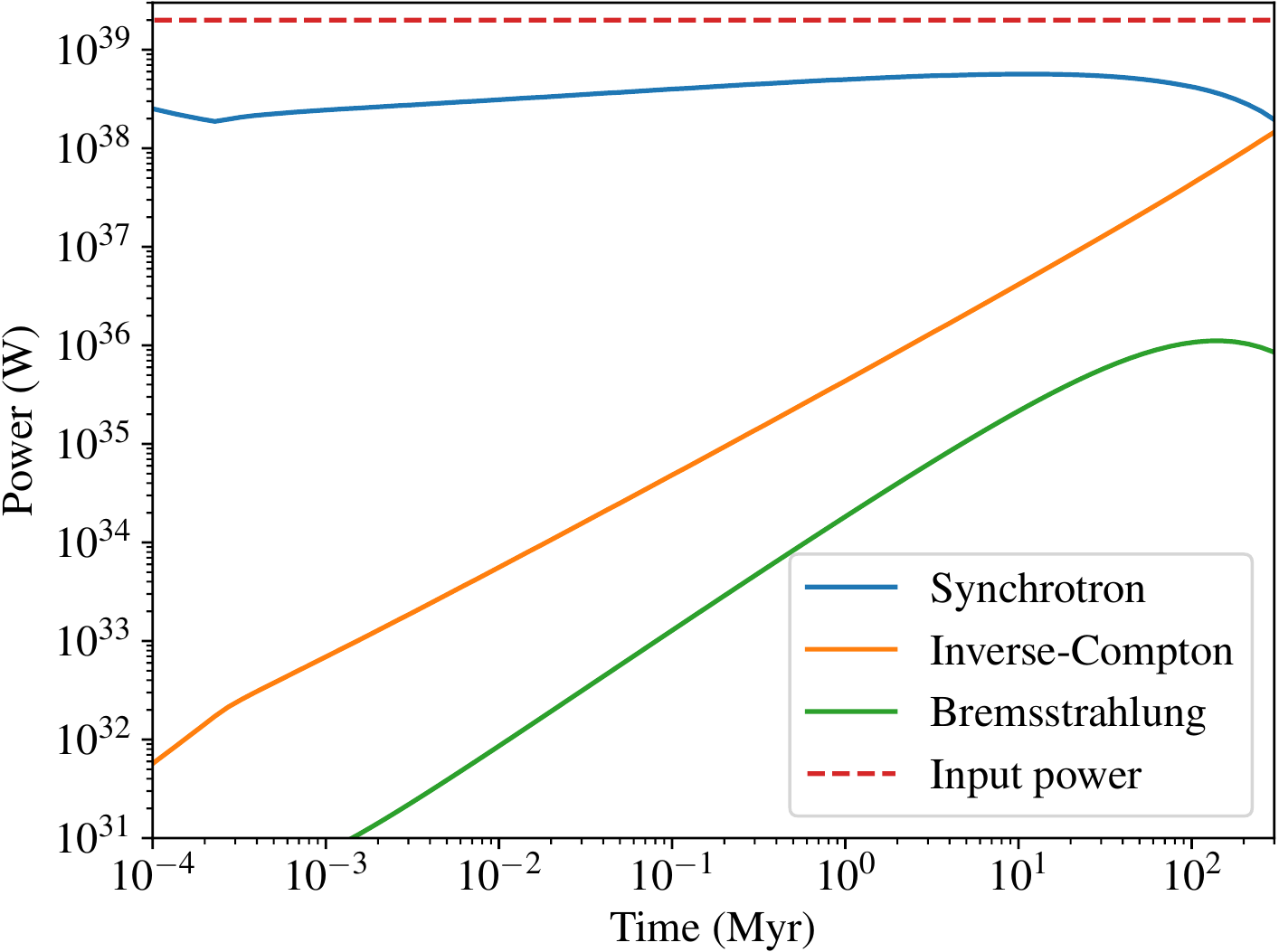}
\caption{Radiative losses compared to the input jet power as a
  function of time for the $Q = 2 \times 10^{39}$ W jet discussed in
  the text. Plotted are the (constant) jet power, the synchrotron
  loss rate from the lobes, the inverse-Compton loss rate from the
  lobes at $z=0$, and the thermal bremsstrahlung loss rate from the
  shocked shell.}
\label{fig:losses}
\end{figure}

I finally carried out a consistency check of assumption (ix), i.e., that
the integrated losses may be neglected compared to the input jet power
in determining the dynamics. To do this it is necessary to integrate
the standard equations for total single-electron synchrotron and
inverse-Compton losses over the electron distribution, e.g. for
synchrotron we have
\begin{equation}
P = \int_{E_{\rm min}}^{E_{\rm max}} \frac{4}{3} \sigma_{\rm T}
  \frac{B^2}{2\mu_0} c \left(\frac{E}{m c^2}\right)^2 N(E) \rd E
\end{equation}
and a similar result holds for inverse-Compton where the energy
density in photons is substituted for the energy density in the field.
Here it is absolutely necessary to apply a correction for losses to
the electron spectrum in the manner described in Section
\ref{sec:loss}, since assuming a power-law spectrum extending up to
high energies greatly overestimates the power $P$. I also roughly
estimated the thermal bremsstrahlung emissivity of the shocked shell
by considering its mean number density, volume and temperature and
applying standard formulas for the integrated loss rate
\citep{Longair10}; this of course does not take into account line
cooling, which will be important for low temperatures. Results for the
$Q = 2 \times 10^{39}$ W jet are shown in Fig.\ \ref{fig:losses}. As
can be seen, the synchrotron losses for this source, which dominate
over the other two processes, are consistently
of the order of 10 per cent of the total jet power; neglecting this
term in considering the dynamics will cause us to overestimate the
volume of the lobes by a modest amount, though of course the
  uncertainties this imposes on the dynamical model are probably
  negligible compared to those on our assumptions on other model
  parameters. The fraction $P/Q$ is a weak
function of $Q$, in the sense that it is lower for lower jet powers.
For this particular source, synchrotron emission is the dominant loss
process except at the very end of the model run, but at high redshift
it is easily possible for the inverse-Compton losses to exceed the
input jet power at late stages of a source's lifetime, and in such a
case the dynamics could well be more significantly affected.
Synchrotron losses would also be significantly higher for values of
$\zeta$ closer to 1. A fully self-consistent treatment of radiative
losses in solving for the dynamics would add significantly to the
complexity of the code and I defer the development of such a model to
a future paper.

\subsection{Source ensembles}
\label{sec:ensemble}

In this section I use the model to generate evolutionary tracks in
terms of observable quantities for active sources. A standard plot in
the study of radio galaxy evolution is the power-linear size
(`$P$-$D$') diagram of \cite{Baldwin82}; here I also consider the
spectral index/linear size (`$\alpha$-$D$') diagram.

\begin{figure*}
  \includegraphics[width=0.48\linewidth]{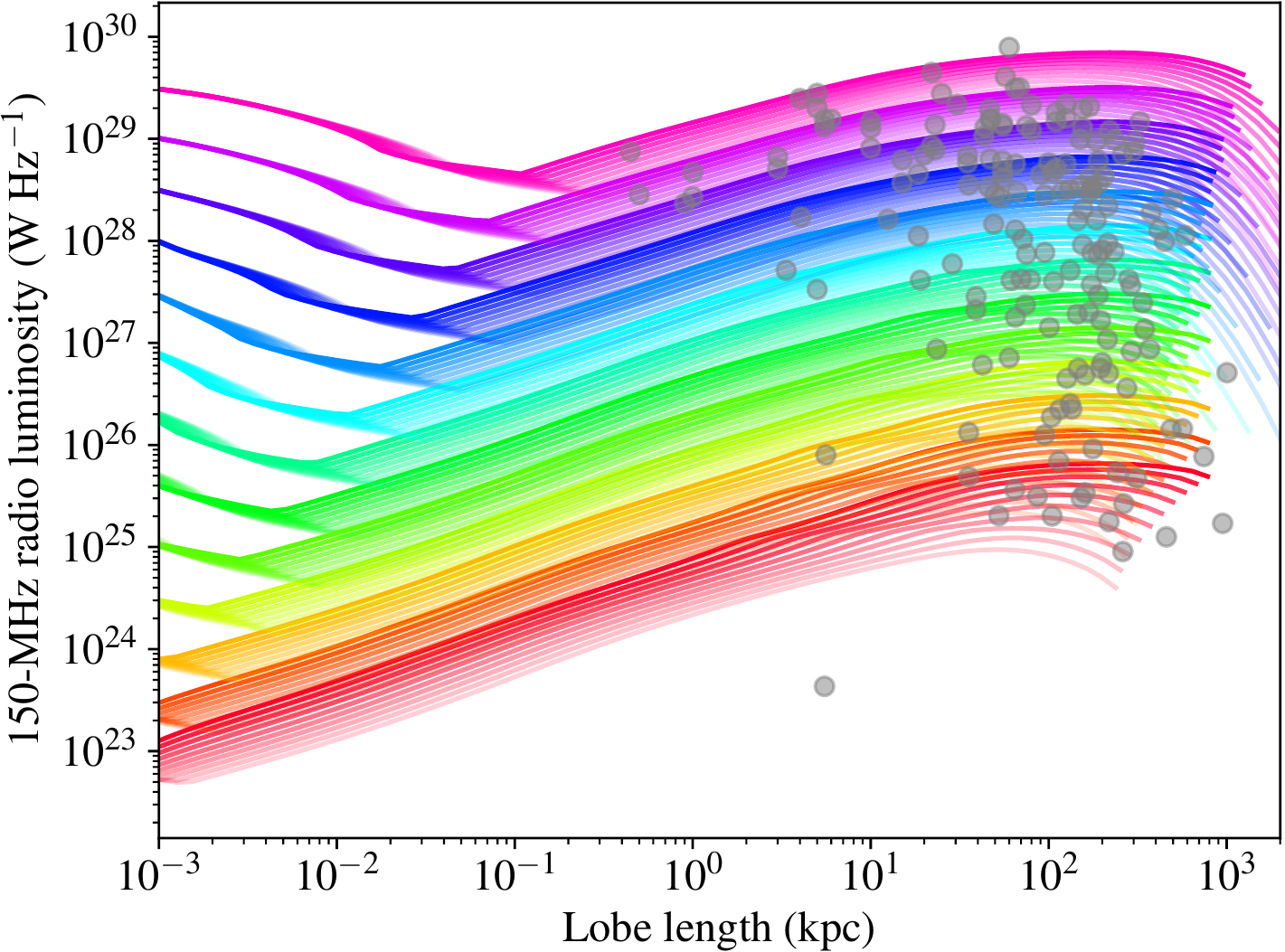}
  \includegraphics[width=0.48\linewidth]{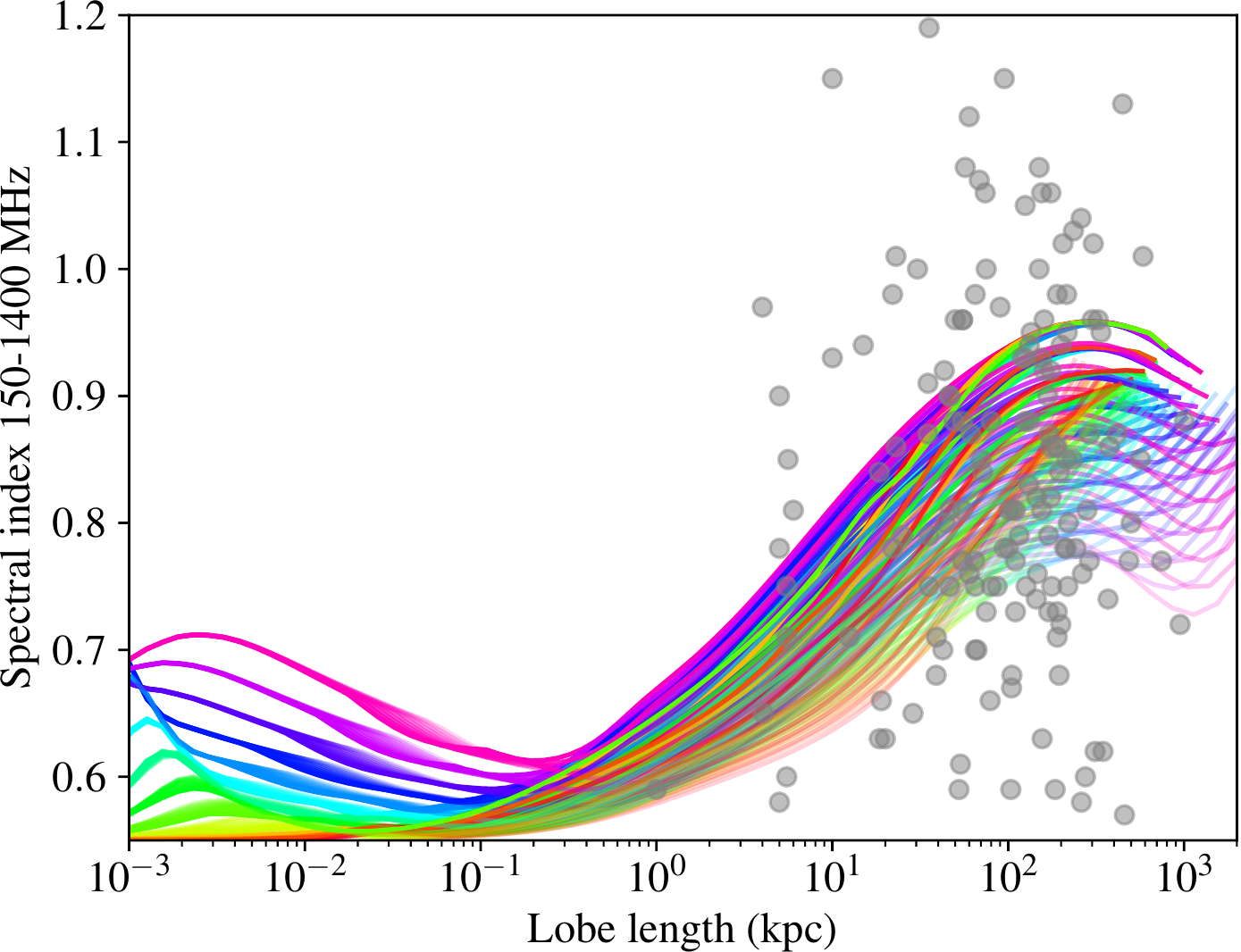}
  \includegraphics[width=0.48\linewidth]{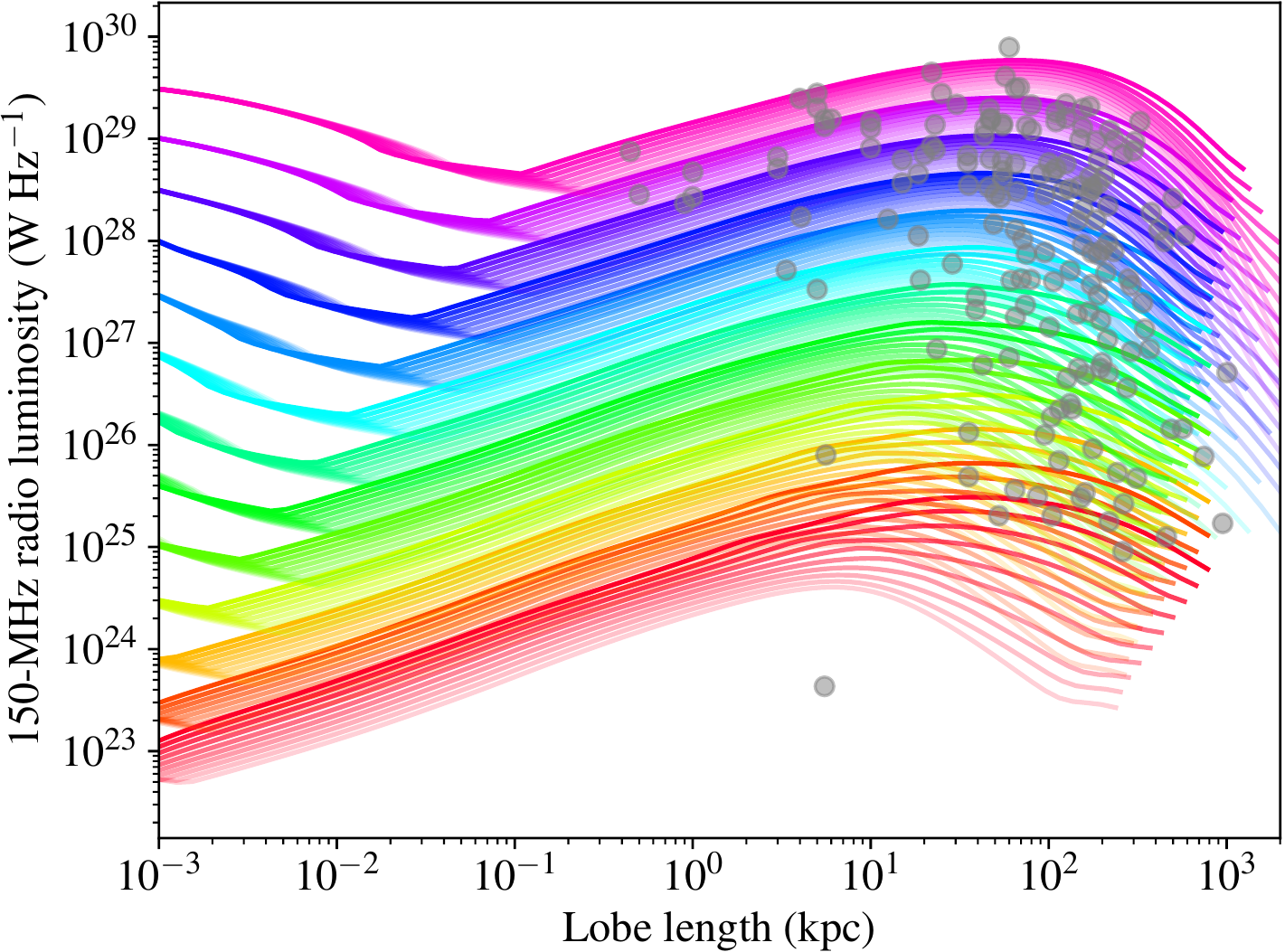}
  \includegraphics[width=0.48\linewidth]{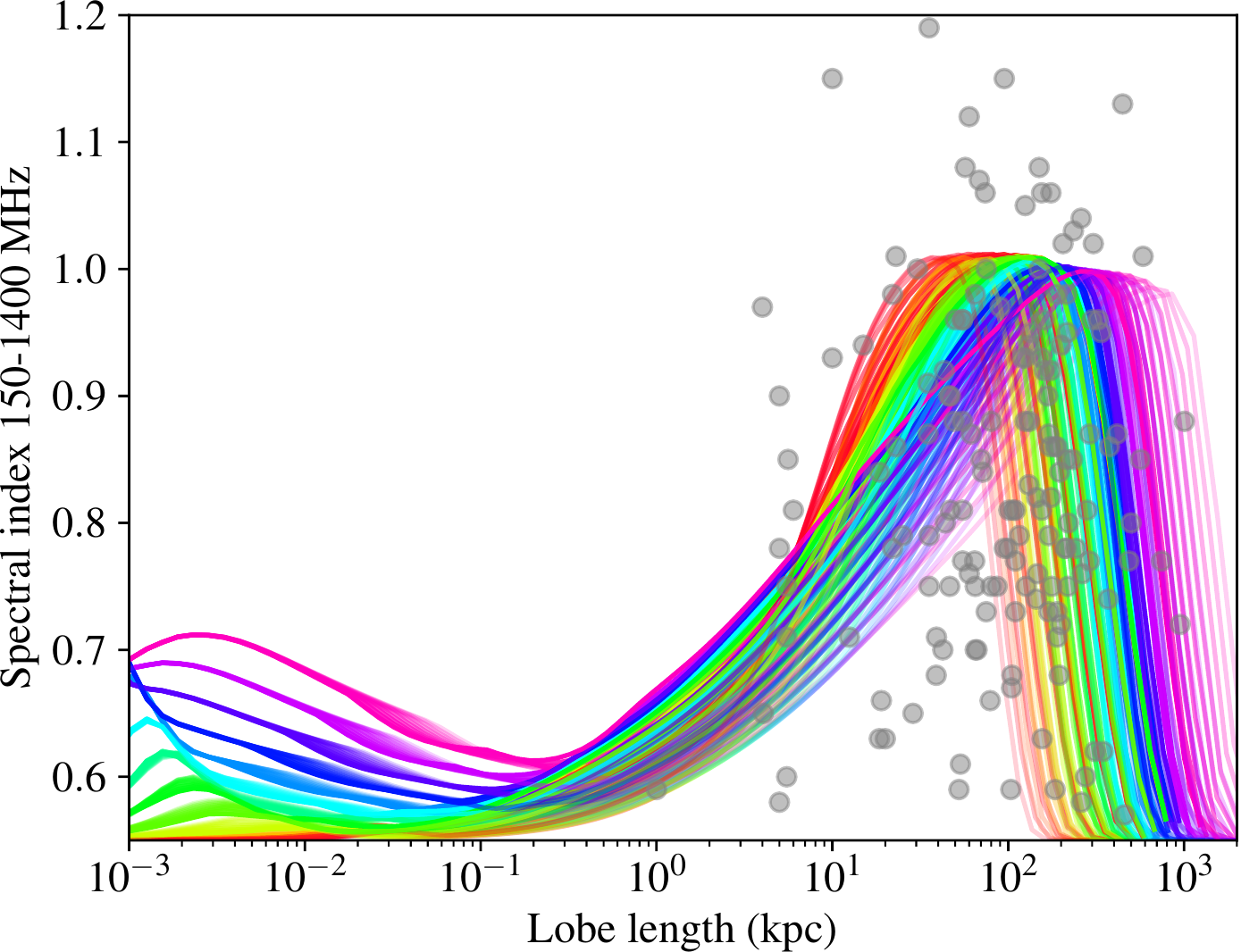}
  \caption{Tracks in the $P$-$D$ and $\alpha$-$D$ diagrams for
    ensembles of sources at $z=0$ (top panels) and $z=2$ (bottom
    panels). All panels show the evolution up to $t=500$ Myr of
    sources with 13 different jet powers evenly spaced in log space
    between $10^{36}$ and $10^{40}$ W; jet power is indicated by the
    rainbow colours (red is low power, violet is high power). Each jet
    power is placed in a universal pressure profile with 10 different
    values of $M_{500}$ between $10^{13}$ and $10^{15} M_\odot$,
    evenly spaced in log space, indicated by the brightness of the
    rainbow colour (faint is low mass, bright is high mass).
    Overplotted on all plots are the rest-frame 150-MHz luminosities
    and the observer-frame 178 MHz--750 MHz spectral indices of the
    3CRR sample, plotted against half their physical (projected) size.
    See the text for more details.}
  \label{fig:pdd}
  \end{figure*}

To populate such a plot I consider sources with jet powers in the
range $10^{36}$ to $10^{40}$ W; I take 13 discrete values of $Q$ in
this range, including the endpoints, evenly spaced in log space. I
consider 10 different universal pressure profile environments with
$M_{500}$ uniformly distributed in log space between $10^{13}$ and
$10^{15} M_\odot$, and allow the sources to evolve in these
environments with constant jet power for 500 Myr (chosen because it
traces even the lowest-power sources out to Mpc sizes; the linear size
axis for the models is truncated, as for some high-power sources in
poor environments the source can grow to scales of tens of Mpc in this
time). As before I assume $\zeta = 0.1$ and $\xi=0.4$.
Fig.\ \ref{fig:pdd} shows the results. Overplotted on this figure are
grey dots representing the 3CRR sources of
\cite{Laing+83}\footnote{Data are taken from the compilation at
  \url{http://3crr.extragalactic.info/}.}. A direct comparison between
the model plots and a sample such as 3CRR is difficult for a number of
reasons. Firstly, 3CRR spans the redshift range from $z=0$ to $z=2$,
and we know from above that the effects of the CMB are important at
high $z$: accordingly I carry out the synchrotron calculation for
both $z=0$ and $z=2$ to show the maximum possible difference.
Secondly, real sources are projected (i.e. their measured physical
size is shorter than their true physical size); for most sources this
will introduce a small systematic error which I do not consider
further here. And thirdly, as I will discuss below, the data
available for these sources is not always of high quality.
Nevertheless some useful conclusions can be drawn.

Looking first at the model sources in the $z=0$ $P$-$D$ plot of
Fig.\ \ref{fig:pdd}, we see that the range in radio luminosity is
mostly driven by the input jet power; swathes of colour representing
the different jet powers are clearly distinguishable across the
diagram. However, up to one order of magnitude dispersion at $z=0$ is
caused by the different source environments by the time the sources
have grown to 100-kpc scales and reached their maximum radio
luminosity. This is most clearly visible on the plot for the
lowest-power sources, but a similar dispersion is present for all jet
powers though it decreases somewhat with jet power. Qualitatively and
quantitatively this dispersion supports the conclusions of
\cite{Hardcastle+Krause13}. Jet power can indeed be inferred crudely
from luminosity alone, but there is a degeneracy between jet power and
environment such that errors of up to an order of magnitude can be
made in doing so, {\it if} there is no relationship in general between
jet power and environment richness, a point that I shall return to
below.

The range of jet powers used here were intended to represent a
plausible range of jet powers for objects in the 3CRR sample, and they
succeed in reproducing the range of radio luminosities very well. We
may conclude, subject to the model assumptions, that the 3CRR jet
power range is more or less the same as the input one, and in
particular that few jets in the Universe exceed powers of $10^{40}$ W.
As such jet powers would imply accretion rates of {\it at least} (i.e.
with 100 per cent efficiency) a few solar masses a year over
timescales of order $10^8$ years, this upper limit appears consistent
with observed black hole masses in the most massive galaxies at the
present day. The main effect of going from $z=0$ to $z=2$ on the
$P$-$D$ diagram is to predict a stronger downturn of the luminosity
with size for large sizes, as I have already noted, and this also
appears to be consistent with the 3CRR observations in that there are
no large objects at the maximum radio luminosity, which would require
much higher jet powers in these models (since the most luminous
objects in 3CRR are also at the highest redshifts). Of course, other
effects, including shorter source lifetimes at high $z$, could be
responsible for this effect. Going to high $z$ also increases the
environmental effect on radio luminosity for a given jet power,
particularly at late times, again because of the additional effects of
the CMB (we do not model any redshift evolution of the universal
pressure profile). For the lowest-power sources the scatter in
luminosity for a given jet power increases to almost two orders of
magnitude in radio luminosity.

The spectral index plots in Fig.\ \ref{fig:pdd} present a more complex
picture and a less good agreement with the observational data. Looking
first at the models, we see that at $z=0$ sources go from
flat-spectrum, $\alpha < 0.7$ on scales $\la 10$ kpc (with lower-power
sources being flatter in spectrum) to steep-spectrum, $0.8 < \alpha <
0.95$, on 100-kpc scales. In the steep-spectrum regime the main driver
appears to be environmental density in the sense that the most massive
environments show the steepest spectra irrespective of jet power. At
$z=2$, we see quite a different picture for the large sources where
spectra are first steep and then flatten again: as noted above this is
because old material is aged out of even the 150-MHz rest-frame band
by inverse-Compton losses at this redshift. The dispersion in spectral
index for a given band is driven mostly by jet power in this regime
(because higher jet powers will imply higher magnetic field strengths
for a given size) but within bands of jet power environment also plays
a role. Roughly speaking the environmental dispersion in $\alpha$ is
of order $\pm 0.05$, so this is a small effect.

The comparison with the 3CRR sources is less good here at least partly
because the available data do not match what I have modelled. What is
plottted for the real sources is the 178-750 MHz observer-frame
spectral index as tabulated by \cite{Laing+83}, whereas what I
calculate, to facilitate comparison with more modern data, is the
rest-frame 150-1400 MHz index. For high-$z$ sources one would expect
the true spectral indices to be steeper than what is calculated from
the model for a given source size and jet power: for low-$z$ sources it might be a little flatter. In
addition, the 750-MHz flux densities used for the calculation
originate from the work of \cite{Kellermann+69} and have
non-negligible error bars. It is beyond the scope of this paper to
derive improved 3CRR integrated spectra, though the data exist to do
so using modern surveys like NVSS \citep{Condon+98} and MSSS
\citep{Heald+15}. Of course, in addition, real radio galaxy
  spectra may be affected by other physical effects, such as
  self-absorption (although that will be irrelevant for most large
  3CRR sources) and relativistic beaming in the core, jets and hotspots. A
  couple of qualitative conclusions may nevertheless be drawn.
  Firstly, the large number of
large sources in 3CRR with flat spectra and comparatively large
physical sizes may be partly explained by inverse-Compton losses at
higher $z$, as shown in the bottom right panel of Fig.\ \ref{fig:pdd}.
Secondly, from the results of Section \ref{sec:remnant}, the
scatter in the integrated spectral index in real sources with respect
to the model prediction, and in particular the number of sources with
spectra steeper than the model can account for, may be partly because
the spectral index is very sensitive to changes in the jet power over
the lifetime of a source -- if we view the behaviour in
Fig.\ \ref{fig:synch-remnant} as the response to a step change in $Q$
then it is clear that smaller decreases in $Q$ could produce smaller
but still significant steepening. It is also possible that variations
in the injection index from the $q=2.1$ assumed throughout this
modelling could be responsible for the steep observed spectra in some
3CRR sources; this effect is not present in our models.

These results on tracks in the $P$-$D$ diagram can be compared to
those of \cite{Kaiser+97}, \cite{Blundell+99} or \cite{Turner+18}; the
most directly comparable earlier plots are in figs 13 and 14 of
\cite{Blundell+99} (note that they use one-sided jet powers, total
source sizes, and radio luminosities in W Hz$^{-1}$ sr$^{-1}$). It is
worth emphasising that all of these comparison models agree in
the broad range of luminosities that are inferred for a given jet
powers and source size, implying good convergence on the basic source
physics despite differences in the detailed modelling. The models of the present
paper, however, show radio luminosities that are still rising at source
sizes $\sim 10$ kpc, whereas in the \cite{Kaiser+97} or
  \cite{Blundell+99} analytical models the luminosity is
monotonically decreasing with time. This is a consequence of the
power-law atmosphere assumption in the earlier work (Section
\ref{sec:previous}).

\subsection{Lobes and plumes}

\begin{figure}
  \includegraphics[width=\linewidth]{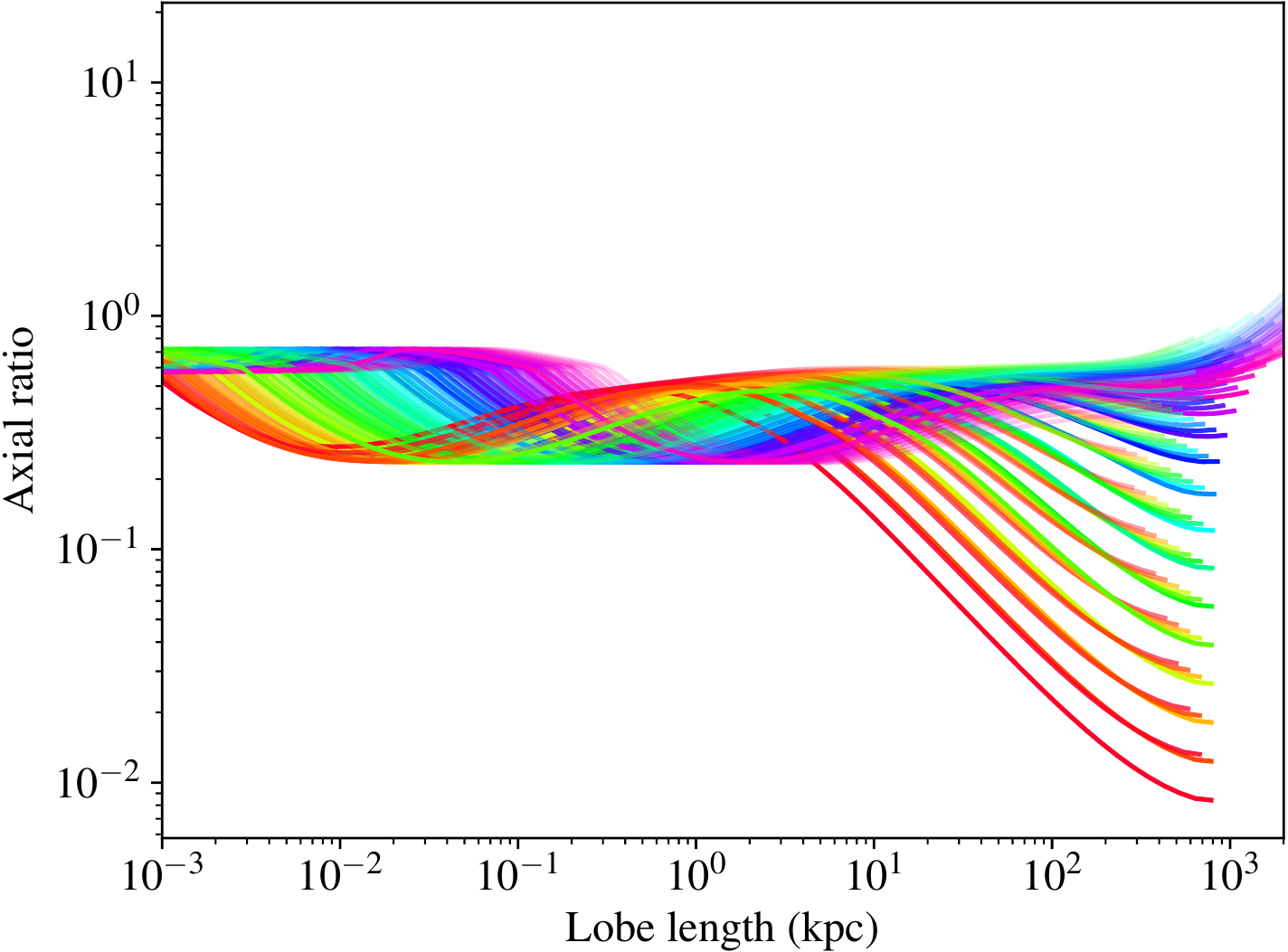}
  \caption{Tracks in the axial ratio/linear size diagram. Colours as
    in Fig.\ \ref{fig:pdd}.}
  \label{fig:lax}
  \end{figure}
  
In Fig.\ \ref{fig:lax} I plot the axial ratio of the sources from the
previous subsection as a function of lobe length, where I define the
axial ratio as the diameter of the lobe divided by its length, and
estimate lobe diameters by taking the volume of the lobe to be a
cylinder with length $R$ and volume given by eq.\ \ref{eq:vl}. Sources
show a characteristic evolution with time, with a constant axial ratio
at early times, then a drop to smaller (thinner) values, then an
increase to a peak value around 0.3-0.4, and then in most but not all
cases a drop again. Sources that are more powerful or in
poorer environments peak later than lower-power sources or those in
richer environments. What is striking about this plot is the existence
of a number of low-power sources in the richest environments that have
low axial ratios (i.e. long, thin lobes). This is simply a consequence
of the slower transverse growth of these sources coupled with the
behaviour of eq.\ \ref{eq:vl} which implies that the energetics of
these low-power, high-density systems is more easily dominated by the
swept-up gas. However, it is certainly qualitatively reminiscent of the
evolution of plumes in FRI sources (including wide-angle tail sources)
in dense environments as discussed in Section \ref{sec:intro:friii}),
and the jet powers involved are consistent with those of such sources,
which at least in some cases are still known to drive shocks on
  large scales \citep[e.g.][]{Simionescu+09}.
Modelling the luminosity evolution of these sources in a way
consistent with the constraints from observation would entail taking
account of entrained material (Section \ref{sec:intro:friii}) and, as
the model does not do this, their positions on Fig.\ \ref{fig:pdd} should be
treated with caution.

\section{Populations and lifetimes}

\subsection{Model setup}

In this section I use the model to consider the effects of
distributions of model parameters and selection effects on observed
source properties.

I generate sources with the basic model parameters of the previous
subsections (in particular $\xi = 0.4$ and $\zeta = 0.1$). It is not
the purpose of this paper to try to match real cosmological
distributions of environment, redshift or jet power -- at this point
we simply do not know these distributions (or, crucially, their
dependences on each other) well enough for this to be a useful
exercise. Instead, I draw sources' redshift from a uniform
distribution over $0\le z<4$, jet powers from a uniform distribution in
log space over $36 \le \log_{10}(Q/{\rm W}) < 40$, and environment from a
uniform distribution in log space such that $13 \le
\log_{10}(M_{500}/M_\odot) < 15$. Sources are also assigned a random
angle to the line of sight distributed such that $p(\theta) =
\sin(\theta)$ for $0<\theta<\pi/2$; the apparent size of a source is
then $R \sin \theta$.

The crucial choice here is the selection of the distribution of source
lifetimes. To model source evolution we need to consider a period in
the immediate past in the frame of each radio source that is long
compared to the typical source lifetime (here we choose 1 Gyr).
Sources are then assumed to be triggered at uniformly distributed
times in that time range -- there is no explicit cosmological
evolution in the model -- so that at the time of `observation' each
source may have switched on with equal probability between 0 and 1000
Myr ago. Every source is evolved according to the model up to the time
of observation. The crucial choice concerns the time period over which
the source has an active jet during the modelling time. Each source is
assigned a lifetime $T$ and is modelled as a remnant (Section
\ref{sec:remnant}) for $t>T$. In what follows I consider two models:
(i) $T$ is drawn from a uniform distribution in linear space between 0
and 500 Myr; and (ii) $T$ is drawn from a uniform distribution in log
space in the range $-3 \le \log_{10}(T/{\rm Myr}) < \log_{10}(500)$.
10,000 sources are simulated for each of these two distributions.

Once the sources' dynamics have been evolved up to the time of
observation I then apply the corrections for radiative and adiabatic
losses described in Section \ref{sec:loss}, computing them only for
the final time step of the simulation. I choose an `observing'
frequency of 150 MHz as the reference frequency and also compute the
corrections needed to calculate {\it observer's-frame} spectral index
between 150 and 1400 MHz. When these calculations are done, many
sources (where the source has been in the remnant phase for a long
time at the time of observation) have 150-MHz luminosities which are
reduced by the loss corrections to negligible values, and hence would
not be detected by any real observation: these sources are considered
`dead'. Sources for which this is not true, and so might in principle
be detected, even if they are remnants, are referred to as `live' in
what follows.

I can then convert intrinsic to observed
quantities using a standard flat $\Lambda$CDM cosmology with $H_0 = 70$ km
s$^{-1}$, $\Omega_{\rm m} = 0.3$ and $\Omega_\Lambda = 0.7$ and
calculate the observed 150-MHz flux density and angular size. This
allows me to mimic observational selection criteria, which give us
both a flux density and a surface brightness limit. To roughly match
bright samples derived from LOFAR surveys (see e.g.\ 
\citealt{Hardcastle+16}) I choose a flux limit of 10 mJy and a
surface brightness limit of a few times the rms noise in LOFAR imaging
of $\sim 100$ $\mu$Jy beam$^{-1}$ with a 6-arcsec Gaussian restoring
beam. Only `live' sources which meet the flux limit and surface
brightness limit are included in the samples I analyse; the intention
is that the selection effects in real data are at least roughly
modelled in the simulations.

Below I discuss some of the results of this modelling for the two
lifetime distributions considered.

\subsection{Basic properties and the effects of lifetime distribution}
\label{sec:lifetime}

Sample (i), where the lifetime distribution is uniform in linear
space, contains 3,453 `live' sources out of 10,000 (34 per cent), of
which 2,112 are selected by the observational criteria. Sample (ii),
where the lifetime distribution is uniform in log space (and therefore
there are many more short-lived sources), contains only 841 live
sources, of which 443 are observable. The mean lifetime (and therefore
the total injected AGN energy) is of course shorter in Sample (ii) and
so no physical conclusions can be drawn from the different source
numbers.

\begin{figure*}
  \includegraphics[width=1.0\linewidth]{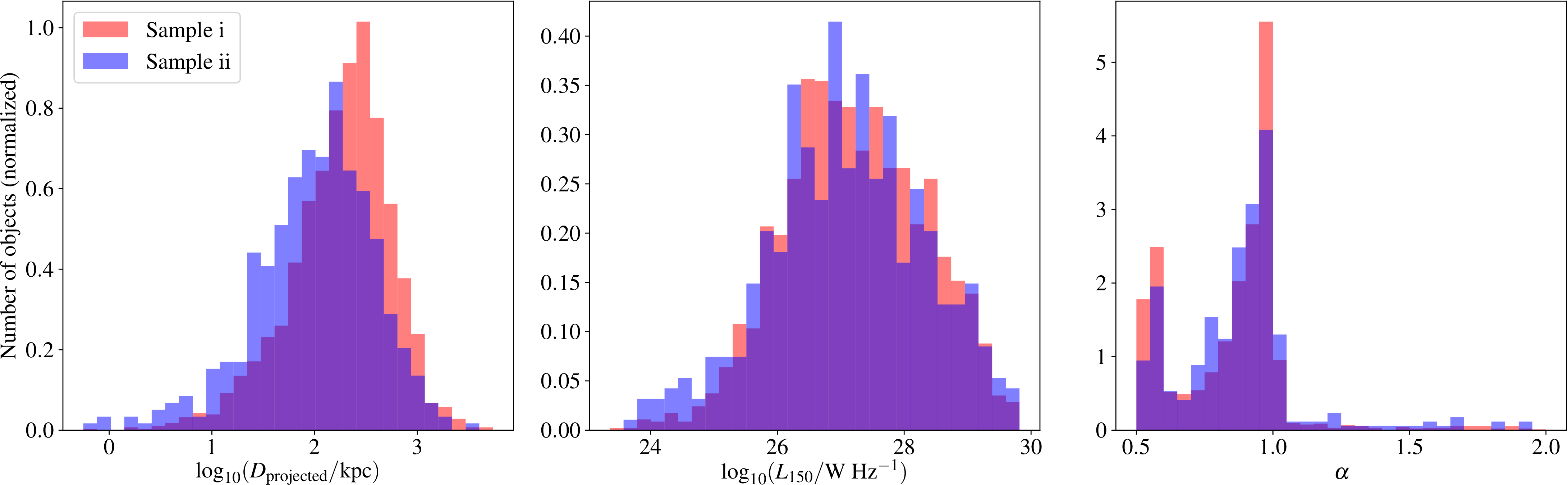}
\caption{Histograms of key observables for the simulated populations.
  The left and centre panels show the full range of the distributions
  in linear size and radio luminosity for the observationally selected
subsamples of Samples (i) and (ii). The range in the right panel has been
chosen to allow the distribution of spectral indices between 150 and
1400 MHz in the
non-remnant population to be seen -- the tail of very steep spectral
indices, associated with extreme remnants, continues to very steep, in
practice unobservable, spectral indices.}
\label{fig:sample-hists}
\end{figure*}

Fig.\ \ref{fig:sample-hists} shows histograms of the distributions of
key observable source parameters for the two samples. We can see that the
two choices for lifetime distributions produce almost identical
distributions of $L_{150}$ and $\alpha$ but clearly distinct source
size distributions -- in the sense that Sample (ii), with typically
shorter lifetimes, produces many more small sources. The median
projected size
in observable sources in Sample (ii) is 117 kpc, whereas it is 206 kpc
in Sample (i). Note that this difference is not nearly as extreme as the
difference in the median lifetimes for the parent samples -- 250 Myr
vs 0.7 Myr -- because most of the short-lifetime sources are low in
luminosity and/or die long before they can be observed. Observational
selection effects favour large, bright sources. Nevertheless this plot
illustrates the importance of the lifetime distribution in determining
the size distribution. If other controlling parameters such as the
environment and jet power distributions were known, the lifetime
distribution could in principle be constrained from observations.

\subsection{Remnants}
\label{sec:ensemble-remnant}

\cite{Godfrey+17} have recently addressed the question of the remnant
fraction expected in sensitive observations. I do not propose to
repeat all of their analysis here, but it is interesting to look at
the effects of our different modelling and selection on the expected
numbers of remnants.

The observed subsamples of Samples (i) and (ii) have slightly
different total remnant fractions, respectively 20 per cent and 34 per
cent (the statistical error on the fraction in Sample (ii) is roughly
3 per cent, so this difference is significant). Again we see that the
lifetime distribution has an effect. It is not hard to see why Sample
(ii), which contains the same number of sources but with shorter
lifetimes, has a higher remnant fraction. Thus the remnant fraction in
principle constrains not just the physics of post-jet source evolution
but also the lifetime distribution.

\begin{figure*}
  \includegraphics[width=1.0\linewidth]{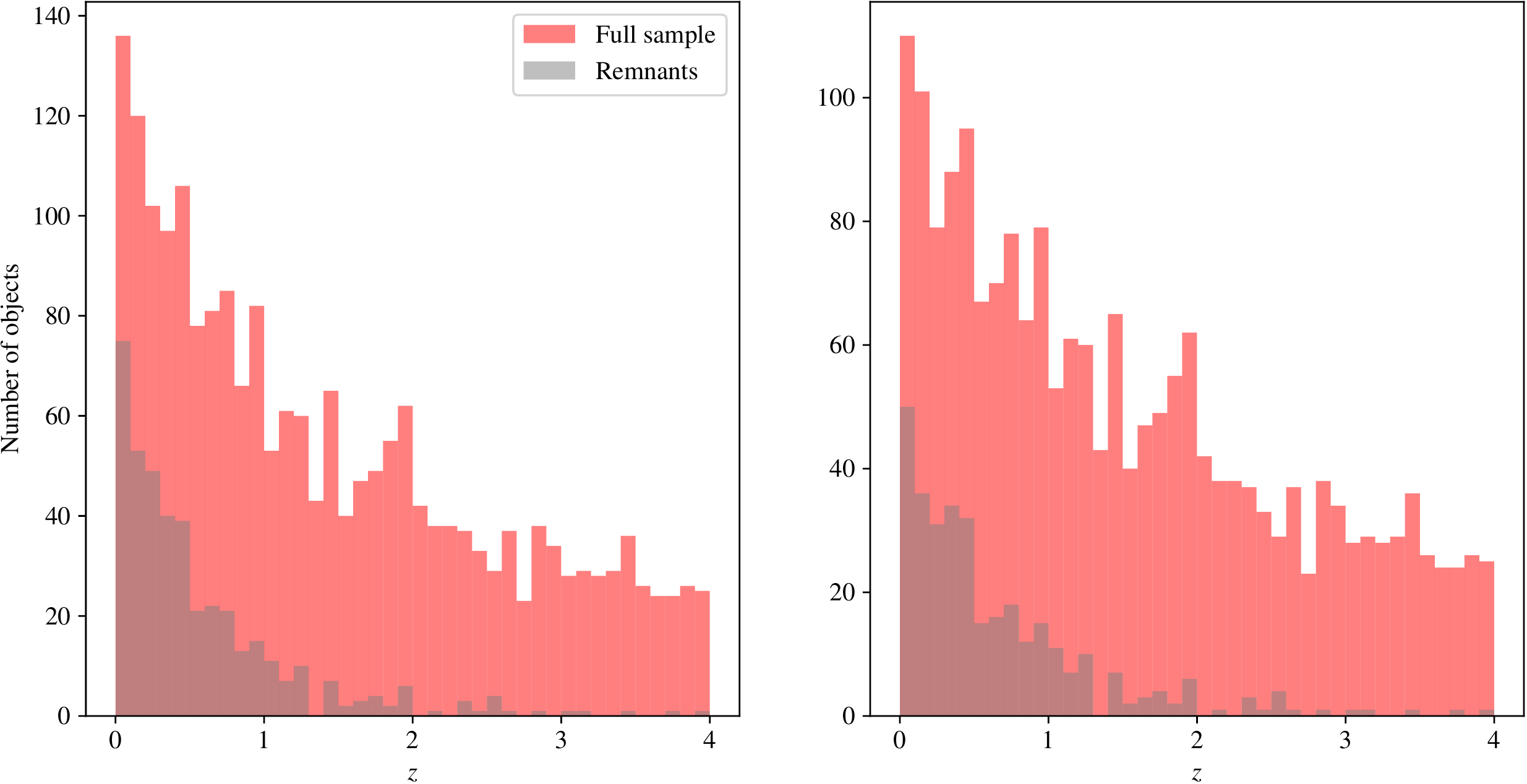}
\caption{The redshift distribution of all sources and of remnant
  sources in Sample (i) as discussed in the text. Left panel: all
  luminosities are shown. Right panel: only sources with $L>3 \times
  10^{25}$ W Hz$^{-1}$ are shown.}
\label{fig:remfrac}
\end{figure*}

Fig.\ \ref{fig:remfrac} shows the remnant fraction as a function of
redshift for Sample (i). (Sample (ii) shows similar trends but with
poorer statistics and we do not consider it further here.) I plot
both the total observed sample, and, for comparison, the distribution
of luminous objects with $L_{150} > 3 \times 10^{25}$ W Hz$^{-1}$, a
luminosity above which the $P$-$D$ diagram is well sampled for active
sources. We can see that the remnant fraction is a strong function of
redshift. This is due to a combination of the faster CMB
inverse-Compton loss rates at high redshift, as discussed above, with
observational selection effects that reduce the numbers of
low-luminosity high-$z$ sources detected. At $z<1$, the fraction of
remnants is 37 per cent (31 per cent above $3 \times 10^{25}$ W
Hz$^{-1}$). It drops to less than 10 per cent for $1<z<2$ and is
basically negligible at $z>2$. Both the remnant fraction and the trend
with redshift are broadly consistent with the results of
\cite{Godfrey+17} (who used a uniform distribution of source lifetimes
up to 200 Myr), despite the many differences in the modelling and
assumptions about source and power distributions, but it is worth
noting that in the \citet{Godfrey+17} models remnants are comparable
to, or even outnumber active sources at low $z$, whereas in the models
of this paper
they never do so. Around 80 per cent of the remnant sources have
`ultra-steep' observer-frame 150-1400 MHz spectral index $\alpha >
1.5$.

\subsection{The radio power/jet power relation}
\label{sec:ensemble-jetpower}
\begin{figure*}
  \includegraphics[width=0.49\linewidth]{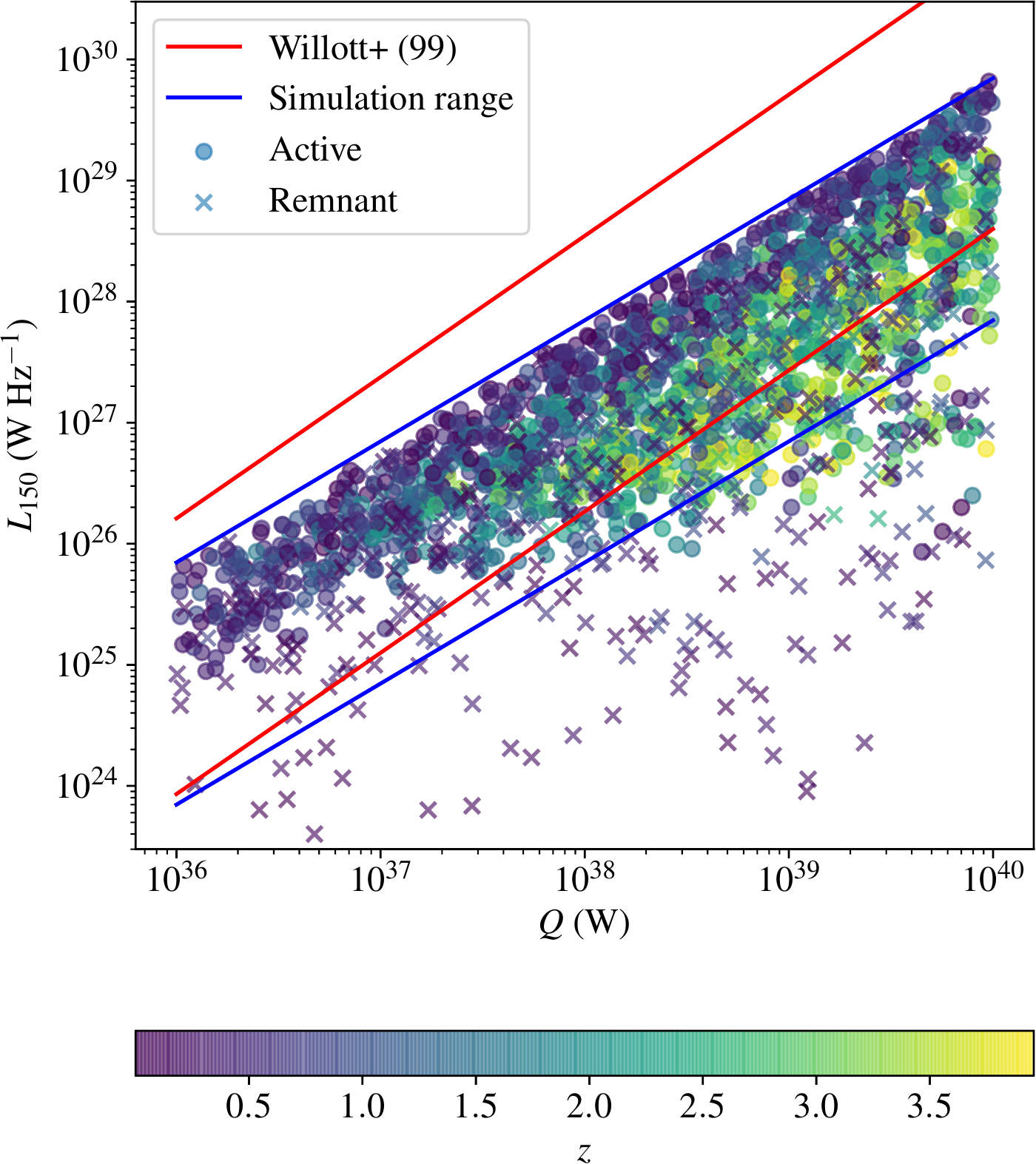}
  \includegraphics[width=0.49\linewidth]{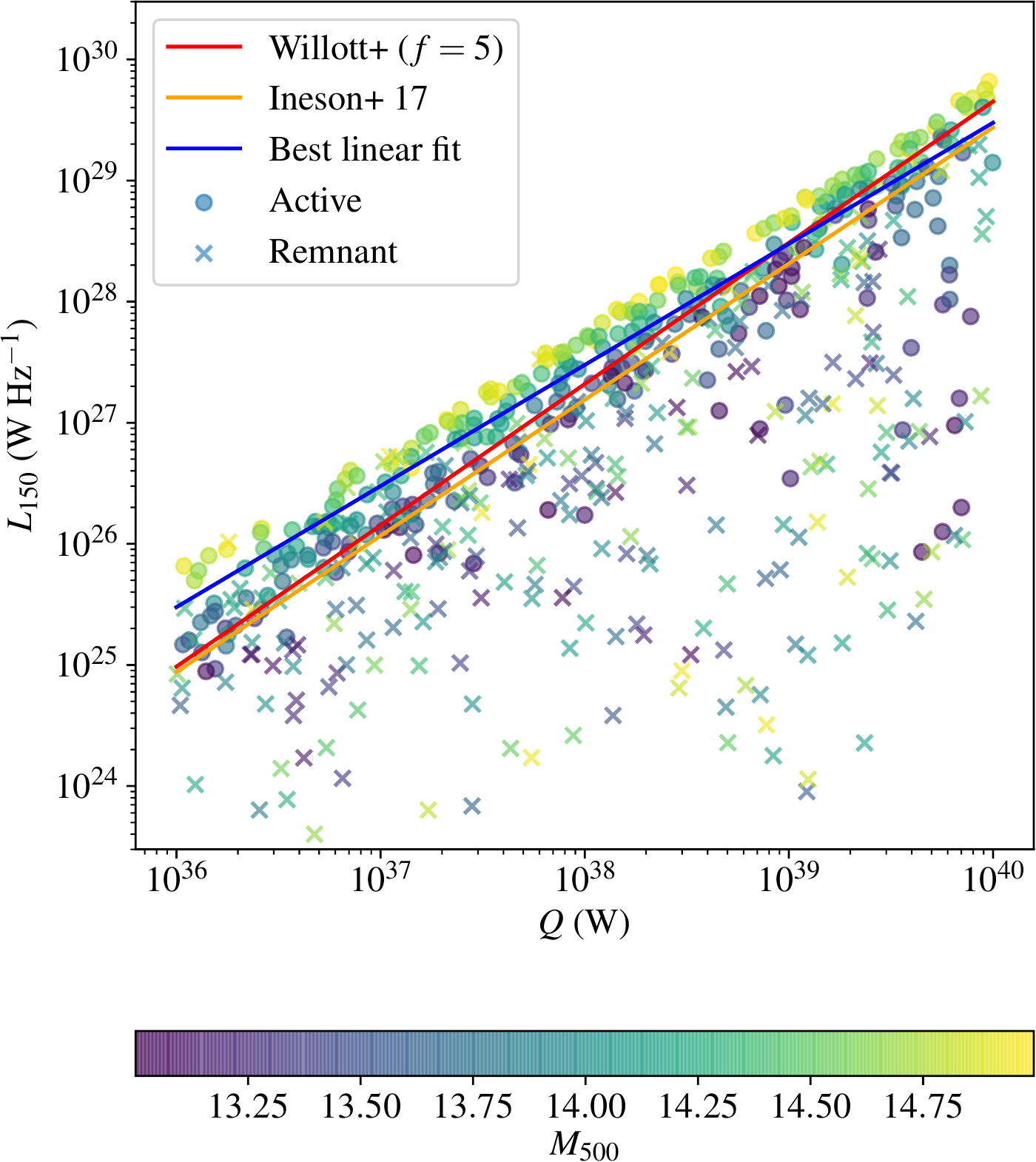}
\caption{Radio luminosity as a function of jet power. Left panel: all
  observed sources from Sample (i), colour coded by redshift.
  Overplotted in blue are lines bounding the relation for active
  sources ($L_{150} = 7 \times 10^{27} (Q/10^{38}\ {\rm W})\ {\rm
    W\ Hz^{-1}}$ and $L_{150} = 7 \times 10^{25} (Q/10^{38}\ {\rm
    W})\ {\rm W\ Hz^{-1}}$) and in red are the \protect\cite{Willott+99}
  relation for $f=1$ (top) and $f=20$ (bottom). Right panel: only the
  $z<0.5$ sources from Sample (i), colour coded by $M_{500}$.
  Overplotted in blue is the best-fitting linear relation, $L_{150} =
  3 \times 10^{27} (Q/10^{38}\ {\rm W})\ {\rm W\ Hz^{-1}}$, in red
  is the \protect\cite{Willott+99} relation with $f=5$, and in orange is the
  relation of \protect\cite{Ineson+17}. }
\label{fig:ql150}
\end{figure*}
  
The modelling allows me to investigate the `observed' radio luminosity
to jet power relation. Here I use exclusively the observed subsample
of Sample (i) since the results should not in this case be dependent
on the lifetime distribution.

Fig.\ \ref{fig:ql150} (left) shows a scatter plot of radio luminosity as a
function of $Q$. Remnants are clear outliers, as expected, but the
non-remnant sources show a good linear correlation, albeit with about
two orders of magnitude scatter (as indicated by lines on the figure)
between the two quantities. Also plotted are the relationships of
\cite{Willott+99} for two values of their factor $f$, $f=1$ and
$f=20$; we see that $f=1$ substantially overpredicts the radio
luminosity for a given jet power but larger values of $f$ do
reasonably well at reproducing the predictions of the model.

The major factor driving the scatter in the relation is redshift, as
can be seen from the colour coding in Fig.\ \ref{fig:ql150}; high-$z$
sources have significantly lower radio luminosity for a given $Q$.
This trend is a result of the increased inverse-Compton losses at
higher redshifts, bearing in mind that observationally selected
sources are biased towards large sizes. If we consider sources in a
narrow redshift band, e.g. as in the right-hand plot of
Fig.\ \ref{fig:ql150}, then the remaining driver of scatter is, as
expected, environment, together with other sources of scatter such as
evolutionary state. For $z<0.5$ there is roughly 0.4 dex (rms) of
scatter about the relation
\begin{equation}
  L_{150} = 3 \times 10^{27} \frac{Q}{10^{38}\ {\rm W}}\ {\rm
    W\ Hz^{-1}}
  \label{eq:jetpower}
\end{equation}
where the normalization is estimated directly from the simulated
observations of non-remnant sources. The \cite{Willott+99} relation
with $f=5$ also fits the data reasonably well. Our best-fitting
relation is also very consistent with the regression line of
\cite{Ineson+17} for their sample of FRIIs with very similar
  jet powers and overall physical properties to those modelled here, and, as they note,
therefore also consistent with observational constraints by
\cite{Daly+12} and with the modelling of \cite{Turner+Shabala15}.

The good agreement between these different observational and
theoretical models is striking and suggests that inference of jet
powers from radio luminosities can be done in a reasonably robust
manner in real samples of FRIIs with known redshifts. However, the
modelling also makes it clear that taking environment and age into account is
important to get a more accurate result -- there is still nearly an
order of magnitude difference between the lower and upper envelopes of
the right-hand plot of Fig.\ \ref{fig:ql150} -- and that special
classes of sources, such as outliers or extreme giants, will be
outliers on any regression. Estimates of the active jet power of
  remnants or even the largest giants naively using the relation of
  eq.\ \ref{eq:jetpower} would lie below the true $Q$ values by several orders of magnitude.

Finally, it is important to note yet again that the radio
  luminosities I calculate in the model depend on model assumptions on
  quantities like $\zeta$ and the jet injection index $q$. If these
  have some intrinsic scatter, or worse still if they are found to
  depend on jet power $Q$ itself in real sources, then there will be
  corresponding dispersion or bias in inferences of jet power using
  the model. The modelling here thus demonstrates the importance of
  constraining the distributions of these quantities and their
  relationship to jet power in large observational samples.

\subsection{Radio luminosity and environment}
\label{sec:rlenv}

\begin{figure}
  \includegraphics[width=\linewidth]{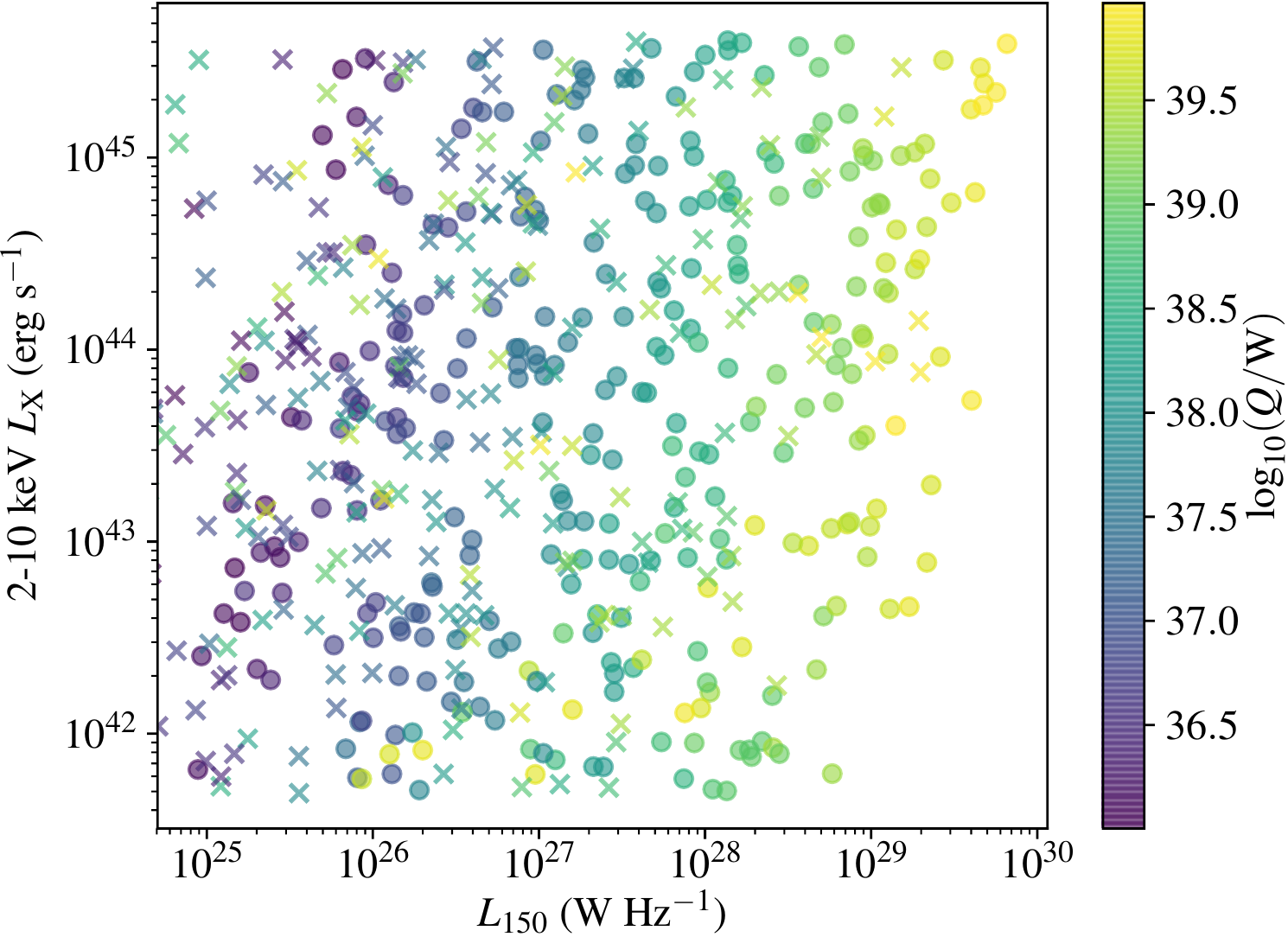}
  \caption{Cluster X-ray luminosity as a function of radio luminosity for $z<0.5$
    objects in the observed subsample (the
    orientation of the plot matches figure 6 of
    \protect\cite{Ineson+15}.) Remnant sources are marked with
    crosses, active sources are filled circles. Sources are
    colour-coded with their jet power so that trends with this
    quantity are visible.}
  \label{fig:m500l150}
\end{figure}

\cite{Ineson+13,Ineson+15} are the latest in a line of authors to
report a correlation between radio luminosity and some measure of
cluster richness -- in their particular case between radio luminosity
and X-ray luminosity of the host environment for the LERG population
only. I searched for any such relation in the simulated observations
of Sample (i) at $z<0.5$, i.e.\ roughly matched to the redshift range
of \cite{Ineson+15}. No such relationship is found in the simulated
data (Fig.\ \ref{fig:m500l150}), where I map $M_{500}$ on to cluster
X-ray luminosity using the relationship of \cite{Pratt+09}, though the
simulated data span exactly the X-ray and radio luminosity range
observed by Ineson et al. For a given jet power, of course, the
cluster mass drives $\sim 1$ order of magnitude of scatter in the
radio luminosity, but most of the variation in radio luminosity is due to
variation in $Q$. This supports the argument of \cite{Ineson+15} that
some intrinsic relationship between the jet power and the cluster mass
might be required to produce such a correlation. In the simulations of
this Section, a
jet of any power can be switched on in an environment of any richness,
and the only observational selection effect is whether it produces a
radio source bright enough to be observed in the sample. In reality,
at least for sources where the hot gas is the source of accreting
material, we might expect an intrinsic $Q$-$M_{500}$ correlation --
and perhaps also a lifetime-$M_{500}$ correlation? -- which
could drive the sort of result seen by \cite{Ineson+15}. This in turn
might be expected to affect the scatter in and the observational slope of jet power/radio
luminosity plots such as those discussed in the previous section.

\subsection{The age-size relation}
\label{sec:agesize}

\begin{figure}
  \includegraphics[width=\linewidth]{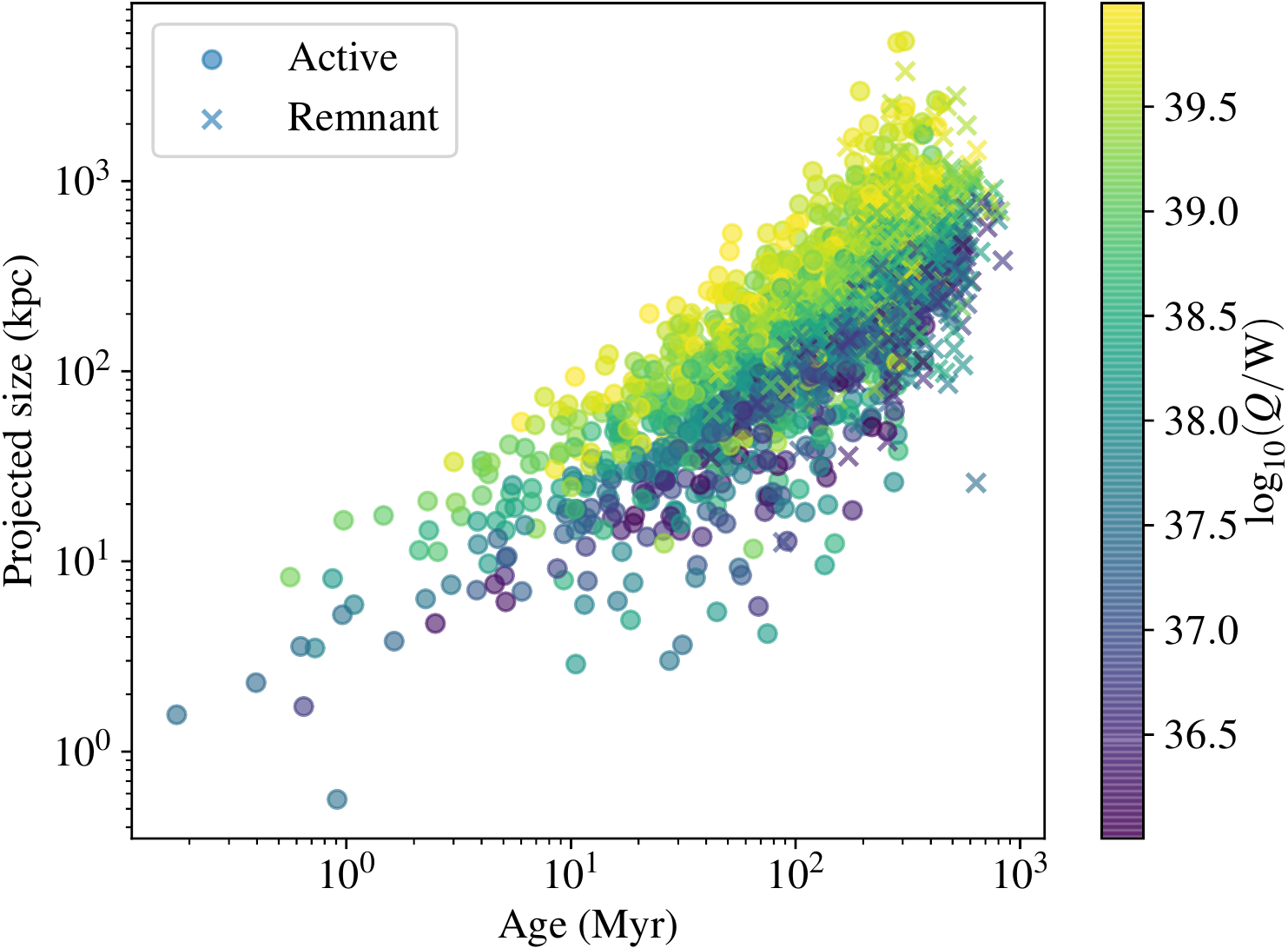}
  \caption{Source linear size as a function of age for the observed
    subsample, colour coded by jet power.}
  \label{fig:agesize}
\end{figure}

The apparent (projected) linear size of a source is expected to be an
indicator of age, but is affected by many other factors, including
projection. Fig.\ \ref{fig:agesize} shows the relationship for the
observed subset of Sample (i). Sources with ages of a few
hundred Myr can have apparent linear sizes of anywhere between a few
and a few hundred kpc. Jet power (as shown on the figure) and
environment affect the position of a source on this plot, but the
unknown projection angle makes the length of any given source a poor
measure of its age. Constraining the projection angle for large
samples will require measurements of the side-to-side depolarization
ratio \citep{Laing88,Garrington+88} which will only be possible for
well-resolved sources (this ratio can be measured in numerical
simulations \citep{Hardcastle+Krause14} but is not currently predicted
by the model of this paper) or of jet sidedness or prominence, which
give more indirect measurements of the angle and again require high
resolution. For sources with the required multiwavelength, resolved
radio images, direct spectral ageing measurements will also be
possible, so it seems unlikely that direct estimates of age from the
size will ever be useful. The most reliable estimate of a source age
in the absence of such resolved imaging might be derived from
estimating the jet power from relationships such as that of Section
\ref{sec:ensemble-jetpower} and then inferring the age from the total
energetic content of the lobes, which depends only weakly on the true
lobe linear size.

\section{Summary and conclusions}

This paper has presented a new `semi-analytical' model of powerful
radio galaxy evolution, which centres around a couple of simple
differential equations to model the evolution of the shock front
around FRII-type lobes, making use of a number of simplifying
assumptions derived from 2D and 3D numerical modelling. In the
  reference models I assume a light, electron-positron jet, no
non-radiating particles in the lobes, and a magnetic field energy
density about a tenth of that in the radiating particles -- all based
on existing observational constraints. Unlike many existing models, I
make no assumptions about self-similar expansion of the lobes or
shocks. These equations are solved numerically, in principle for a
general spherically symmetrical environment, in practice for
isothermal $\beta$ models or the `universal pressure profile' of
\cite{Arnaud+10}; the radio luminosity and spectral evolution of the
simulated sources can then be determined in post-processing. I show
good agreement between the dynamics of the model and earlier numerical
simulation work, as expected since the model's assumptions are based
on numerical simulations. The model can reproduce the broad features
of the evolution of the synchrotron luminosity also seen in numerical
models. The evolution of the integrated spectrum of a source from flat
to steep, as seen observationally, is also reproduced. {\sc python}
code to carry out all of these steps is publicly available at
\url{https://github.com/mhardcastle/analytic}.

Key results of the paper are as follows:
\begin{itemize}
  \item The typical integrated
    spectrum of an aged source is smoothly curved over two decades of
    radio frequency, rather than being a broken power law (Section
      \ref{sec:results_loss}).
  \item The relatively low magnetic field
    strengths in the fiducial models imply an important role for
    inverse-Compton losses at high $z$. Thus the radio luminosity
    evolves more strongly with time, and is in general lower for a
    given jet power, at high redshift than at $z=0$; this qualitative
    behaviour agrees with earlier work, but the quantitative effect
    will depend on assumptions about magnetic field strength (Section
      \ref{sec:results_loss}).
  \item Remnant sources, after the jet has switched off, show a rapid
    evolution of radio luminosity and spectral index due to the
    cessation of injection of new flat-spectrum electrons combined
    with radiative and adiabatic losses of the previously existing
    population (Section \ref{sec:remnant}). This is consistent with recent work by
    \cite{Godfrey+17}.
  \item In early stages of source evolution a steep spectrum for a
    given lobe length is a good marker of a rich environment. However,
    radio galaxies in general, and high-$z$ sources in particular, can
    move to a low-luminosity, flat-spectrum state at late times where
    the most recently injected electrons dominate the
    integrated radio emission. Thus selection of either high-$z$ radio
    sources or sources in rich environments by their spectral indices
    needs to be carried out with care (Section
      \ref{sec:results_loss},\ref{sec:ensemble}).
  \item If jet power and environment are unrelated, which may not be
    the case in real objects, jet power is the main driver of radio luminosity variations.
    The plausible two orders of magnitude variation in cluster mass
    that I model, however, gives rise to one to two orders of
    magnitude variation in radio luminosity for a given jet power and
    redshift (Section \ref{sec:ensemble}). Thus simple inference of the jet power from radio
    luminosity is unsafe.
  \item Studies of simulated sources with different lifetime
    distributions shows that the distribution of source lifetimes has
    a significant effect on both the source length distribution and
    the fraction of remnant sources expected in observations
    (Section \ref{sec:lifetime}). The
    remnant fraction is expected to be low ($\sim 30$ per cent) even
    at low redshift and low observing frequency due to the rapid
    luminosity evolution of remnants, and to tend rapidly to zero at
    high redshift due to inverse-Compton losses (Section \ref{sec:ensemble-remnant}).
  \item Simulated observations reproduce a strong correlation between
    low-frequency radio luminosity and jet power (Section
    \ref{sec:ensemble-jetpower}), which is in excellent agreement with
    recent observational work, but only a poor correlation between
    source apparent size and age (Section \ref{sec:agesize}).
    Source redshift, age and environment all also affect the
    luminosity.
  \item No radio luminosity/environment correlation is expected in
    samples in which the jet power and environment are independent .
    The observation of such a correlation in real data suggests a physical
    relationship between jet power and hot-gas environment, perhaps
    mediated by the source of fuel for accretion and/or by black hole
    mass (Section \ref{sec:rlenv}).
\end{itemize}

Although the model described here is encouragingly consistent
  with both observations and numerical models of larger samples, it is
  clear that its adjustable parameters should be refined by further detailed
  study of small numbers of sources in the regime that it best
  describes. In a forthcoming paper Mahatma \etal\ will compare the
  model predictions with radio and X-ray observations of two powerful
  cluster-centre FRII sources. At a later stage we hope to apply it to large
samples of powerful sources generated from, for example, LOFAR surveys
observations \citep{Hardcastle+16,Shimwell+17} with the aim of
inferring the kinetic luminosity function of powerful jets and hence
the work being done by radio-loud AGN on their environments at the
present day.
    
\section*{Acknowledgments}

I am grateful to Judith Croston, Martin Krause and Judith Ineson for helpful
discussion of many of the topics covered in this paper, and to Vijay
Mahatma, Martin Krause and Judith Croston for helpful comments on
earlier drafts. I thank an anonymous referee for constructive comments.

I acknowledge support from the UK Science and Technology Facilities
Council [ST/M001008/1]. This research has made use of the University
of Hertfordshire high-performance computing facility
(\url{http://stri-cluster.herts.ac.uk/}). This research made use of
{\sc Astropy}, a community-developed core Python package for astronomy
\citep{AstropyCollaboration13} hosted at
\url{http://www.astropy.org/}, and of {\sc topcat} \citep{Taylor05}.

\bibliographystyle{mnras}
\renewcommand{\refname}{REFERENCES}
\bibliography{../bib/mjh,../bib/cards}

\end{document}